\documentstyle[11pt,psfig]{article}
\begin{document}
\renewcommand{\thepage}{ }
\begin{titlepage}
\title{
\hfill
\vspace{.5cm}
{\center \bf 
Phase diagram of doped spin-Peierls systems
}
}
\author{
Michele Fabrizio$^{(a,b,c)}$, R\'egis M\'elin$^{(d)}$, and
Jean Souletie$^{(d)}$\\
{}\\
{$^{(a)}$ International School for Advanced Studies SISSA-ISAS,}\\
{Via Beirut 2-4, 34013 Trieste, Italy}\\
{}\\
{$^{(b)}$ Istituto Nazionale di Fisica della Materia INFM}\\
{}\\
{$^{(c)}$  International Centre for Theoretical Physics,}\\
{Strada Costiera 11, 34014 Trieste, Italy}\\
{}\\
{$^{(d)}$ Centre de Recherches sur les Tr\`es Basses
Temp\'eratures (CRTBT) \thanks{U.P.R. 5001 du CNRS,
Laboratoire conventionn\'e avec l'Universit\'e Joseph Fourier
}}\\
{CNRS BP 166X, 38042 Grenoble Cedex, France}\\
}
\date{February 1999}
\maketitle
\begin{abstract}
The phase diagram of a model describing doped CuGeO$_3$ is derived. 
The model emphasizes the role of local 
moments released by the impurities and  
randomly distributed inside the gaped singlet background. 
The phase diagram is investigated by two
methods: (i) in a mean field treatment of the interchain
coupling and (ii) in a real space decimation procedure in a  
two dimensional model of randomly distributed moments. 
Both methods lead to similar results, 
in a qualitative agreement with
experiments. In particular, a transition to an inhomogeneous 
N\'eel phase is obtained for arbitrary small doping. From
the decimation procedure,  
we interpret this phase at very low doping as a 
{\sl Griffith antiferromagnet}. Namely, it does not
have a true long range order down to zero temperature.
Nonetheless, large magnetically ordered 
clusters appear already at relatively high temperatures. 
This demonstrates the role of disorder in the theoretical
description of doping in CuGeO$_3$.
A detailed comparison with other approaches is also given.
\end{abstract}
\end{titlepage}
\newpage
\renewcommand{\thepage}{\arabic{page}}
\setcounter{page}{1}
\baselineskip=17pt plus 0.2pt minus 0.1pt
%%%%%%%%%%%%%%%%%%%%%%%%%%%%%%%%%%%%%%%%%%%%%%%%%%%%%%%%%%%%%%%%%%%

\newpage
\section{Introduction}

Spin-Peierls systems have been the subject of a renewed interest
because of the discovery of several inorganic quasi-one dimensional
(Q1D) spin-Peierls compounds.
A 
very well studied example is CuGeO$_3$. This Q1D compound made of
weakly coupled CuO$_2$ chains has a 
spin-Peierls transition at $T_{SP}\simeq 14K$ \cite{SP}, 
as revealed by 
diffraction measurements, showing a dimerization below $T_{SP}$, and by 
susceptibility data, which indicate the opening of a spin gap at 
$T_{SP}$. 

CuGeO$_3$ can be doped in a controlled 
fashion, giving the unique opportunity to study the role of
impurities in a 
spin-Peierls system. Doping is done in two different ways. 
One possibility is to directly affect the spin degrees of freedom 
within each CuO$_2$ chain by substituting some of the Cu atoms
(whose spin is $S=1/2$) with
magnetic (Ni \cite{Ni}, which has $S=1$, or  
Co \cite{Co}, $S=3/2$) or non-magnetic (Zn \cite{Zn,Hase,Martin}, Mg \cite{Mg}) 
impurities. 
Another way is by substituting Ge with Si \cite{Si}, which is 
expected to modify the strength of the exchange couplings among the magnetic 
ions. For all these compounds, even a small amount of doping strongly 
enhances the magnetic fluctuations, leading in most cases to a lower 
temperature N\'eel phase coexisting with the Peierls's distorted phase. 
For instance, in 
0.3\% Si-doped compounds, free spins are released which 
contribute with a Curie-like behavior below $T_{SP}\simeq 12.5K$ (which is 
slightly reduced with respect to the undoped sample), and finally 
get frozen below the N\'eel temperature $T_{N}\simeq 0.95K$ \cite{Grenier}.  
The same behavior has been reported in a $2 \%$ Zn-doped compound,
where magnetization measurements can be interpreted as if approximately one
spin-1/2 per Zn atom is released  \cite{St-Paul}.
In all cases but Co-doped samples \cite{Cobalt}, 
the low doping $x<1\%$ phase 
diagram seems to have universal features \cite{Grenier2}:
$T_{SP}$ decreases almost linearly with $x$, 
while no critical concentration for the appearance of the N\'eel phase 
has till now been revealed down to the lowest accessible impurity 
concentrations (e.g. 0.12\% Zn for 
which $T_N\simeq 0.025K$ \cite{Znnew}).
  
At larger doping, the situation is still not fully established. However, 
most of the data indicate that over some critical $x_c$,
the spin-Peierls long-range order disappears, although short-range 
correlations seem to persist. For Mg-doped samples, it is claimed that at 
$x_c\simeq 2.3\%$ a first order phase transition takes place
which is signaled by a  jump of $T_N$ \cite{Mg}.

Manabe {\it et al.}  \cite{Znnew} have reported an
antiferromagnetic (AF) 
ordering with Zn concentrations as low as 0.11\%. Their analysis of
the variation of the N\'eel temperature versus doping concentration
has shown the absence of a critical doping concentration
for AF ordering at zero temperature ($x_c=0$).
The goal of the present article is to understand
this onset of long range AF order at such low doping.
This experimental observation questions the validity
of existing theories in a much more drastic way than
the coexistence of AF order and dimerization.

Several distinct theories describe this coexistence of dimerization
and antiferromagnetism. Fukuyama {\it et al.}
 \cite{Fuku} first proposed a model showing that the two
orders can coexist, and another model was
proposed by Mostovoy {\it et al.}  \cite{Mostovoy}.
Two of the present authors have proposed another model
for the coexistence of AF ordering and dimerization
\cite{usuno,usdue}. All these theories successfully describe
the coexistence
of dimerization and AF order upon doping, and they also
show that the effect of doping amounts
to generate states inside the Spin-Peierls gap, in
agreement with susceptibility experiments (for
example  \cite{Znnew}). A similar behavior is reported in spin-ladder 
systems, which, in spite of having a large spin gap,  
undergo a N\'eel transition upon very small doping. For instance 
SrCu$_2$O$_3$, with a spin gap of $\sim 400K$, has a finite $T_N\simeq 3K$ 
already at $1\%$ Zn doping \cite{Azuma}. In these ladder compounds the 
low energy spin excitations induced by doping have also been
revealed  by a finite specific heat coefficient \cite{Azuma}.

In their article, Manabe {\it et al.}
pointed out that one energy scale  emerging from our
model (denoted by $T_*$, see sections
\ref{sec:sus} and \ref{sec:Mean-Field} for
its definition) was many orders
of magnitude lower than the N\'eel temperature (for instance
at $x=0.1 \%$, $T_* \simeq 4 \times 10^{-42}$K whereas
$T_N \simeq 20$mK has been measured \cite{Znnew}).
On the basis of the analysis of AF fluctuations in the
1D model, we already suggested \cite{usuno,usdue} that this
apparent inconsistence can be avoided if one makes
a proper treatment of disorder averaging. In the present
article, we push further this analysis by estimating
the N\'eel temperature in the presence of interchain coupling,
which turns the quasi-long-range AF fluctuations of the 1D
model into a longer range order.
We prove by two different methods (a mean field treatment
of the interchain coupling, and a decimation procedure
of the 2D disordered problem) that the
resulting N\'eel temperature has the right order of magnitude.
Moreover, under both treatments, we show that there is no
critical concentration for the establishment of AF order.
As it was shown by Mostovoy {\it et al.}, this is
not the case in their model, and we believe that this
is not the case either in the approach by Fukuyama
{\it et al.}, further extended by Yoshioka and Suzumura
\cite{Suzumura}.

The paper is organized as follows. 
We introduce the model in section \ref{sec:themodel}, and
discuss the experiments by Manabe {\it et al.} \cite{Znnew}.
The problem of interchain coupling is first solved in section
\ref{sec:Mean-Field} at the mean field level.
Section \ref{sec:renor} is devoted to the decimation
approach that goes beyond the mean field treatment of
interchain interactions, and is more relevant for
the extremely low doping situation. The picture
of Griffith-like antiferromagnetism emerges from this treatment:
large AF correlated clusters are formed when the temperature
is decreased below the spin-Peierls temperature, without
infinite range ordering.
A summary of our results is presented in section
\ref{sec:conclusion} together with a comparison
with the approaches by Mostovoy {\it et al.} \cite{Mostovoy},
Fukuyama {\it et al.} \cite{Fuku}, and Yoshioka and Suzumura
\cite{Suzumura}.
We argue that these approaches
would fail to properly describe the low doping behavior, whereas
our description leads to AF ordering at infinitesimal doping.
Since the analysis of the work by Fukuyama {\it et al.}
is more technically involved, we have kept the qualitative
features for the concluding section while the technical
aspects are left for the appendices. 

\section{The Model}
\label{sec:themodel}
\subsection{Generation of spin-1/2 moments}
\label{sec:moments}
We start by briefly describing our model. Let us consider 
the Hamiltonian of a dimerized Heisenberg model. 
Given the large anisotropy in the couplings
(for instance \cite{neutrons})
\begin{equation}
\label{eq:parametres}
|J_a| \simeq  J_c / 100 \ll J_b \simeq J_c / 10 \ll J_c
\simeq 10.6 \mbox{meV}
,
\end{equation}
where $c$ is the chain axis, and $a$ and $b$ the two
perpendicular directions,  
the N\'eel temperature 
is mainly determined by $J_c$ and $J_b$. 
We therefore  consider   
the Hamiltonian of a dimerized Heisenberg model in two dimensions,
\begin{equation}
J\sum_{n,i} \left(1+\delta (1)^{i+n}\right) \vec{S}_{n,i}\cdot\vec{S}_{n,i+1} +
J_\perp \sum_{n,i}  \vec{S}_{n,i}\cdot\vec{S}_{n+1,i}, 
\label{HHeis}
\end{equation}
where $J=J_c$, $J_\perp=J_b$, $i$ labels the sites along the chain and $n$ 
is the chain index. $\delta$ is the dimerization parameter 
which is staggered along the $b$ axis. 
In reality, frustration 
has to be added to get a reasonable agreement with the spin susceptibility 
data above $T_{SP}$ \cite{frustration}.
This can be accomplished by introducing for instance 
a (quite large) next nearest neighbor exchange $J'\sim 0.23-0.36 J$, 
which can result both from next nearest neighbor Cu-Cu superexchange 
and from the lattice quantum fluctuations. However, this additional
complication is neglected in our present analysis.
  
\begin{figure}
\centerline{\psfig{file=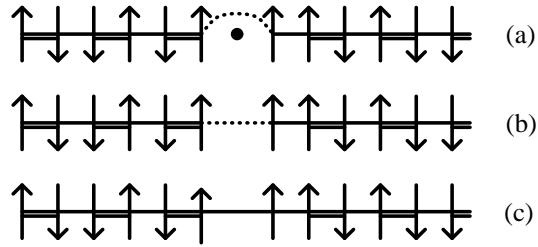,width=7cm}}
\caption{Mapping of a segment with a non magnetic site to a 
squeezed segment, without this site. The double line represents a strong 
bond in the dimerized chain, i.e. $J(1+\delta)$, the single solid line 
a weak bond $J(1-\delta)$, while the dotted line is the weak link across 
the non magnetic site. Going from (b) to (c) we have assumed  
the weak link equal to the weak bond. Therefore the squeezed 
segment in (c) contains a domain wall, {\it i.e.} two consecutive 
weak links.  
}
\label{mapping}
\end{figure}

When $\delta>0$, each spin at 
an even site is coupled by a strong bond $J(1+\delta)$ to the spin at 
its left, and by a weak bond $J(1-\delta)$ to the spin at its right, both 
on the same chain. A finite gap between the singlet ground state and the 
lowest triplet excitation arises as soon as $\delta\not =0$ and 
$J_\perp$ is less than a critical value which depends on $\delta$. 
This should be the case of the pure CuGeO$_3$.  
We start by assuming that the most important effect of doping is that 
each impurity releases a free spin out of the 
dimerized singlet background. This behavior was anticipated
in \cite{Dagotto1} on the basis of exact diagonalization
of small-size systems, and in Ref. \cite{usuno} by more qualitative arguments
(see Fig. \ref{mapping}). 
%
%%%%%%%%%%%%%%%%%%%%%%% FIGURE %%%%%%%%%%%%%%%%%%%%%%%%%%%%%%%%%
%%%%%%%%%%%%%%%%%%%%%%%FIGURE END %%%%%%%%%%%%%%%%%%%%%%%%%%%%%%
%
Further numerical works have confirmed
this picture \cite{Dagotto2,Dagotto3},
and have shown its robustness when more realistic assumptions
are made, such as quantum phonons or coupling to a 3D
lattice dimerization \cite{Poilblanc1},
or next-nearest-neighbor frustration \cite{Poilblanc2}.
The generation of spin-1/2 objects   
is most clearly seen for
Zn or Mg doping and in the limit $J_\perp=0$.   
As a first approximation, the non magnetic ion  
cuts the chain into two segments, e.g. one starting with a weak bond, the 
other ending with a strong bond. In the former segment,  
a free spin-1/2 moment will appear localized at the boundary. 
Even if we add a weak link across the non-magnetic ion, which couples 
the two end-spins of each segment, this free spin-1/2 moment still 
exists. An even more transparent explanation is obtained by considering a ring
of even number of sites. 
Since $\delta \not = 0$, the ground state is a singlet with 
a gap to the lowest $S=1$ excitations. If a spin is removed, and a weak link 
across the empty site is added, the effective spin chain will now contain 
an odd number of spins 1/2. Hence the ground state will have $S=1/2$, 
in spite of a finite gap to higher spin states. 
Moreover, since translational symmetry is broken, this $S=1/2$ moment  
will be mostly localized around the weak link.      
Similarly, even if one spin 1/2 is substituted by a spin 1, modeling 
Ni doping, a free spin 1/2 moment is formed around the impurity. 
We also believe that the same situation holds with a finite $J_\perp$.

The case of Si-doping is less clear. Ge is supposed to   
influence the superexchange between the Cu's \cite{Braden}, increasing 
the antiferromagnetic superexchange with respect to the ferromagnetic 
indirect exchange, finally leading to a net $J>0$. 
The smaller bond length of Si-O has likely the effect of reducing 
the antiferromagnetic superexchange by diminishing the angle 
Cu-O-Cu, thus inducing defects in the dimerized structure. In principle, 
a large reduction of that angle might even change the sign of the 
Cu-Cu exchange constant, leading to the formation of free spin-1 
moments. Although we believe this is likely to occur, 
here we will limit our analysis to the simpler case of non magnetic 
ions substituting Cu.

We therefore take for granted the appearance of spin-1/2 moments 
upon doping, in the line of several numerical works 
(see e.g. Refs. \cite{Dagotto1,Dagotto2,Dagotto3,Poilblanc1,Poilblanc2})
and pursue the next step which is to build up the
phase diagram of doped CuGeO$_3$ from these excitations. 
This cannot be done without including the effect of interactions
between those spin-1/2 moments, which are central to our
description.

As we argued, in such cases free spin-1/2 moments form around 
each impurity, showing up as low energy states inside the spin-Peierls gap, 
which, at very low doping, is almost unchanged. Therefore, we
model the doped compounds as a two components system, 
where spins 1/2 are diluted inside 
a singlet background, characterized by a fixed gap between the singlet ground 
state and the first excited triplets.   
The virtual polarization of the singlet background 
provides an antiferromagnetic coupling 
between these spins. Since the dimerized state is gaped, this coupling 
decays exponentially with the distance between two spins over a length 
of the order of the spin-Peierls correlation length. 
For instance, let us consider two spins   
at a distance $\vec{r}=(n_x c,n_y b)$, 
$n_x$ and $n_y$ being positive integers, and $c$ and $b$ the lattice constants 
in the chain and perpendicular to the chain directions. Then,   
the exchange between these two impurity-released spins 
can be taken of the form
\begin{equation}
J(\vec{r}) \simeq -(-1)^{n_x+n_y}\Delta 
\exp{\left(- \sqrt{\left(\frac{n_x }{\xi_{x}}\right)^2 +
\left(\frac{n_y }{\xi_{y}}\right)^2 } \right)},
\label{J}
\end{equation}
where $\Delta$ is the spin gap,
$\xi_x = \xi_{SP} \simeq 9 - 13$
 \cite{soliton} 
the correlation length in lattice constant units 
along the c-axis, while $\xi_{y} \sim 
\xi_{SP}J_\perp/J \simeq 0.1 \xi_{SP}$, the correlation length 
along the b-axis. The oscillating factor in front takes into account the 
fact that the singlet background polarizes
mostly in a staggered way, leading to a 
ferromagnetic exchange if the two spins are on the same sublattice and 
to an antiferromagnetic one in the opposite case.  
Since the impurities are located at random, replacing a fraction $x$ of the 
magnetic sites, Eq.(\ref{J}) defines a two dimensional random Heisenberg 
model, which should properly describe the excitations at 
energies below the spin-Peierls gap. 

The above two-component model looses its validity at larger doping, where 
the impurities start to affect the singlet background in a non negligible way.
Nevertheless, we believe that at low doping it does  
capture the essential physics of the real system, giving a good 
description of the the N\'eel transition. 
Below some temperature, these weakly coupled spins will tend to order 
antiferromagnetically. Moreover, they will also induce a staggered 
polarization on the singlet background, which does have a finite 
staggered susceptibility, in spite of an exponentially decaying 
uniform susceptibility. According to this scenario, the N\'eel phase   
would consist of spins antiferromagnetically ordered 
diluted in a partially polarized singlet background. The main problem is to 
get an estimate of the N\'eel transition temperature, which
allows a direct comparison of the present model with the experimental phase
diagram of Cu$_{1-x}$Zn$_x$GeO$_3$ \cite{Znnew}, where it
has been shown that zero temperature AF ordering occurs
at an infinitesimal Zn concentration.

\subsection{The typical energy scale $T_*$ and the susceptibility
experiments by Manabe {\it et al.} \protect \cite{Znnew}}
\label{sec:sus}

We first proceed by considering the extremely low doping susceptibility
experiments by Manabe {\it et al.} \cite{Znnew}, which,
to our knowledge are the only available with Zn
concentrations as low as $1.12 \times 10^{-3}$.
As pointed out in Ref. \cite{Znnew},
the N\'eel temperatures (see table \ref{table}) are indeed
much larger than the {\it typical energy scale} $T_*$, 
given by the N\'eel
temperature of a regular array of equally spaced spin-1/2 moments
with an exchange given by Eq. (\ref{J}), which we may estimate
by considering, instead of a random distribution, a regular
distribution of the 
spins over the 2D lattice with a concentration $x$. 
The coupling along the $c$ and $b$-axes are then 
$J_x = \Delta \exp(-l/\xi_x)$ and $J_y = \Delta\exp(-l/\xi_y)$, respectively, 
with $l=1/\sqrt{x}$ the average distance between the spins.
Hence, at low doping, the effective Heisenberg model describing the spins 
has a huge space anisotropy $J_x/J_y \sim \exp(-l/\xi_y)\ll 1$, 
implying that the scale controlling the N\'eel
temperature is $T_* \sim J_y \sim 2 \times 10^{-14}$K if
$x=0.1 \%$, 
which is far too small compared with the 
experimental values of the AF transition temperatures. 
The typical energy scale $T_*$ can also be evaluated in the single
chain version of our model, as recalled in section \ref{sec:Mean-Field}.
Experimentally, Manabe {\it et al.} \cite{Znnew} measure
$0.027$K$< T_N < 0.7$K in CuGeO$_3$, for concentration
impurities ranging from $0.11 \%$ and $0.49 \%$ (see Table
\ref{table}). A more accurate estimate of the
characteristic exchange interaction can be made by fitting their
data in the range $T_N < T < T_{SP}$.

%%%%%%%%%%%%%%%%%%%%%%% FIGURE %%%%%%%%%%%%%%%%%%%%%%%%%%%%%%%%%
\begin{figure}
\centerline{\psfig{file=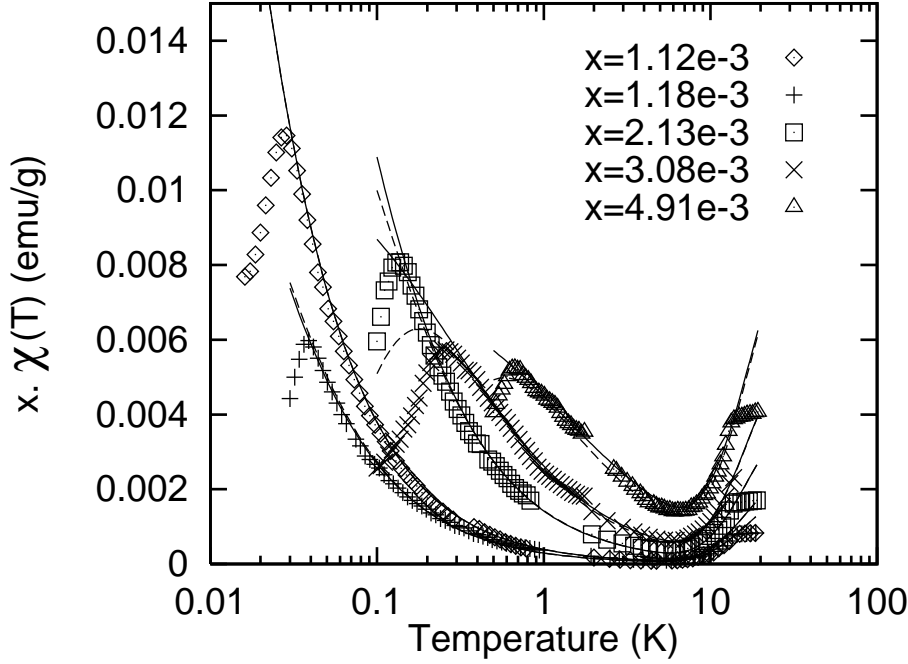,height=8cm}}
\caption{Fit of the susceptibility experiments by Manabe
{\it et al.} \protect \cite{Znnew} to the Curie form
(\ref{eq:curie}) (solid lines) and to the form
(\ref{eq:fit}) (dashed lines).
For clarity, we have mutiplied the susceptibility
by the concentration, so that all the concentration
data are at the same scale at low temperature.
Since the data provided
to us by Manabe {\it et al.} were very dense in
some temperature regions, we have ``compressed''
the data by averaging them over given bins.
The parameters of the
fits are shown on Table \protect \ref{table}.
}
\label{fig:curie}
\end{figure}
%%%%%%%%%%%%%%%%%%%%%%%FIGURE END %%%%%%%%%%%%%%%%%%%%%%%%%%%%%%

%%%%%%%%%%%%%%%%%%%%%%% TABLE  %%%%%%%%%%%%%%%%%%%%%%%%%%%%%%%%%
\begin{table}
\begin{center}
\begin{tabular}{|@{}c@{}||@{}c@{}|@{}c@{}||@{}c@{}||@{}c@{}|@{}c@{}||@{}c@{}||@{}c@{}|@{}c@{}|}
\cline{1-9}
{ }$x$ { }&{ } $J_{\rm imp}$  { }&
{ }$C_{\rm imp}$  { }& 
{ }$C_{\rm SP}$  { }& { } $\theta$ { } &  
{ } $C$ { } & { } $T_N$ { } &
{ } $T_*$ { } & $<<J>>$\\
($\%$) & (K) & { } ($10^{-5}$ eum/g) { } &
{ } ($10^{-5}$ eum/g) { }& (K) & 
{ } ($10^{-5}$ eum/g) { }& (K) & (K) & (K) \\
\cline{1-9} \cline{1-9}
0.112 (2) & 0.0033 & 0.35 & 297 & 0.0035 & 0.35 & 0.027 & 
$10^{-42}$ & .45 \\
\cline{1-9}
0.118 (2) & 0.008 & 0.25 & 215  & 0.01 & 0.25 &  0.04 &
$10^{-39}$ & .47 \\
\cline{1-9}
0.213 (4) & 0.04  & 0.7  & 242  & 0.035 & 0.69 & 0.14 &
$10^{-21}$ & .84 \\
\cline{1-9}
0.308 (6) & 0.18   & 1.0  & 242 & 0.29 & 1.1 &  0.24 &
$10^{-14}$ & 1.2\\
\cline{1-9}
0.491 (10) & 0.58   & 1.6 &  229 & 1.34 & 2.14 & 0.7 &
$10^{-8}$& 1.9\\
\cline{1-9}
\end{tabular}
\caption{Parameters of the best fits of the susceptibility
data by Manabe {\it et al.} to the forms
(\ref{eq:fit})} and (\ref{eq:curie}) shown on Fig. \protect
\ref{fig:curie}. The N\'eel temperature is also given,
and is of the same order of magnitude as $J_{\rm imp}$
and $\theta$. The values of $<< J >>$ (Eq. (\ref{eq:averageJ})) and
$T_*$ (Eq. (\ref{T*})) are also given.
\label{table}
\end{center}
\end{table}
%%%%%%%%%%%%%%%%%%%%%%% END OF TABLE %%%%%%%%%%%%%%%%%%%%%%

We first decompose the susceptibility in the whole
range between $T_N$ and $T_{SP}$
into two contributions:
\begin{equation}
\label{eq:fit}
\chi(T) = \frac{C_{\rm imp}}{T} \exp{\left( - \frac{J_{\rm imp}}
{T} \right)}
+ \frac{C_{\rm SP}}{T} \exp{\left( - \frac{\Delta}
{T} \right)}
,
\end{equation}
the first term corresponding to the low temperature contribution
of Zn impurities, and the second term describing the higher
temperature behavior of the susceptibility
close to the spin-Peierls temperature. The energy
scale involved in this contribution is the spin-Peierls
gap $\Delta \simeq 44.7 \, \mbox{K}$.
The fits for various impurity concentrations
are shown on Fig. \ref{fig:curie},
and are in a quite satisfactory agreement with experiments.

The three parameters $C_{\rm imp}$, $J_{\rm imp}$,
$C_{\rm SP}$ are shown on Table \ref{table}.
The energy scale $J_{\rm imp}$ involved in the fit of the
impurity contribution sets the characteristic
energy scale involved in the
physics of impurity-released magnetic moments. 
As it can be seen
on Table \ref{table}, $J_{\rm imp}$ is finite, which shows that
the Zn-induced magnetic moments are interacting, consistent with
our picture. Moreover, this energy scale is
{\it not} of the order of the typical exchange $T_*$.

Alternatively, the susceptibility data in \cite{Znnew}
can be fitted with the Curie-Weiss expression
\begin{equation}
\label{eq:curie}
\chi(T) = \frac{C}{\theta + T}
+ \frac{C_{\rm SP}}{T} \exp{\left( - \frac{\Delta}
{T} \right)}
,
\end{equation}
a fit proposed in Ref. \cite{Grenier}. 
As it is visible
on Fig. \ref{fig:curie}, the form of the susceptibility
also leads to a good description of the susceptibility
for temperatures ranging between the N\'eel temperature
and the spin-Peierls temperature. The parameters of the fits
are shown on Table \ref{table}, and  the energy scale
$\theta$ involved in the fit (\ref{eq:curie})
is again orders of magnitude larger than the typical exchange $T_*$.
Both expressions
(\ref{eq:fit}) and (\ref{eq:curie}) have the same high
temperature behavior in the limit $T \gg J,\theta$,
and, in this regime, $\theta = J$. Once the two fits
have been performed independently, we can check that
these energy scales are indeed the same, as it is visible
on table \ref{table}, and comparable to the ordering
temperature $T_N$.

The reason why the N\'eel temperature is not of the order of
$T_*$, but many orders of magnitude higher, is
that, as a result of disorder in the spatial distribution
of magnetic moments, clusters exist in the system with spins
which are close to each other, being therefore strongly 
AF correlated. Once the interchain couplings will be included
(sections \ref{sec:Mean-Field} and \ref{sec:renor}), this will result in
a N\'eel temperature many orders of magnitude larger
than $T_*$.

\section{Mean field theory of AF ordering}
\label{sec:Mean-Field}

In this section we first recall (section \ref{sec:1D})
the results obtained in
1D \cite{usuno,usdue}, discuss the energy scale $T_*$ in
the 1D model (section \ref{sec:T*}), 
and present in section \ref{sec:Jperp}
the extension to
include the effect of the interchain coupling $J_{\perp}$.
We take in this section as well as in the forthcoming section
\ref{sec:renor} a spin-Peierls gap $\Delta = 44.7$K
\cite{neutrons},
assumed to be independent on temperature.
The spin-Peierls
correlation length is $\xi_{SP}=9$ \cite{soliton}.

\subsection{Physics of the 1D model}
\label{sec:1D}
As shown in \cite{usuno,usdue}, 
a single chain does develop quasi-long range magnetic correlations 
at zero temperature, which we believe is the key point
for understanding the establishment of AF ordering at
arbitrarily low doping. More precisely, the correlation length 
$\xi(T)$ is found as the
root of $f(y,\Gamma_T) + 1 = 0$, with $y=-1/\xi$,
$\Gamma_T = \ln{(\Delta/T)}$, and
$$
f(y,\Gamma) = \alpha
\frac{f_0(y) - \alpha \tan{(\alpha \Gamma)}}
{f_0(y) \tan{(\alpha \Gamma)} + \alpha}
,
$$
where $\alpha = \xi_{SP} \sqrt{-y(y+2 x)}$, and
$f_0(y)= \xi_{SP}(y+x)$. The same solution holds in
the XX limit, except that the correlation length is
then found as the root of the equation $f(y,\Gamma_T) + 2 =0$.
The correlation length is compared on Fig. \ref{fig-corre} to
that of an ordered array of
magnetic moments separated by a distance of $1/x$
and interacting with an exchange $\Delta \exp{(-1/x \xi_{SP})}$.

Finally, we will use latter the local staggered
susceptibility $\chi_s(T)$
\begin{equation}
\label{eq:chis}
\chi_s(T) = x \left( \frac{1/\xi_{SP}}
{x \ln{(\Delta/T)} + 1/\xi_{SP}} \right)^2 \frac{1}{4 T}
.
\end{equation}

\subsection{Energy scale $T_*$}
\label{sec:T*}
\begin{figure}
\centerline{\psfig{file=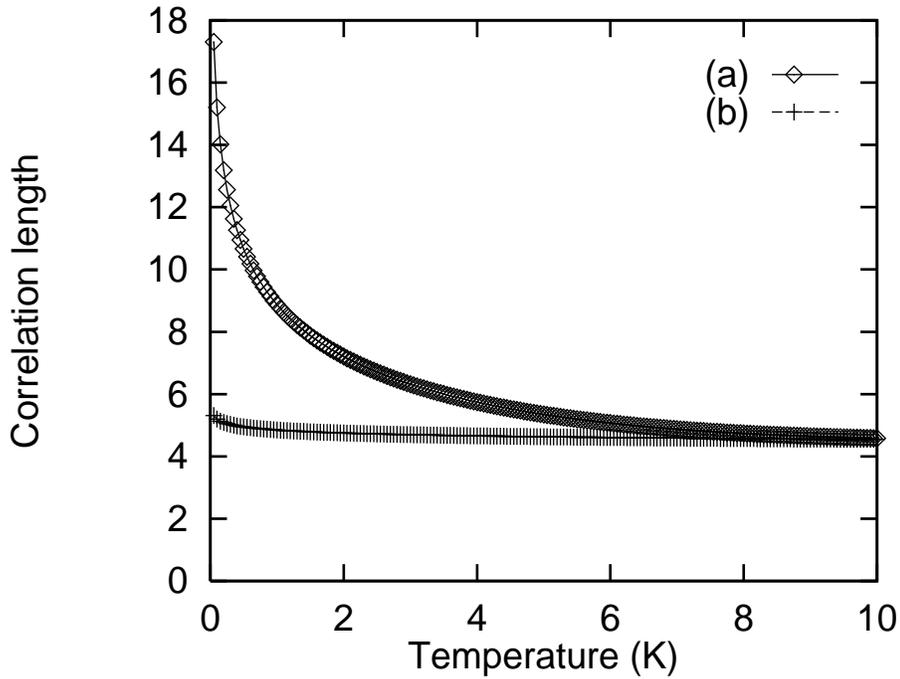,height=8cm}}
\caption{Variations of the correlation length
in the XX limit 
for $x=0.3 \%$ doping versus temperature, with
$\xi_{SP}= 9$ and $\Delta = 44.7\,\mbox{K}$.
(a) shows the variations of the correlation length
of the disordered system and (b) the correlation
length of the typical realization of disorder,
namely, an array of equally spaced magnetic moments.
The limiting value of the correlation length
when $T \rightarrow T_{SP}$ is 
$1/(n+2/\xi_{SP})$, the factor of $2$ originating
from the XX approximation.
}
\label{fig-corre}
\end{figure}

The staggered spin-spin correlation 
function
\[
\langle S^i(r)S^i(0) \rangle \sim (-1)^r \frac{{\rm e}^{-r/\xi(T)}}{r^2},
\]
shows quasi-long-range AF fluctuations at low temperature, with
the limiting behavior of the correlation length
\begin{equation}
\label{eq:xiT}
\xi(T)\to \frac{2 x \xi_{SP}^2}{\pi^2} \ln^2\left(\frac{T}{\Delta}\right)
\end{equation}
valid at extremely low temperature $ T< T_*$, in which case the system
has crossed-over to the random singlet fixed point \cite{usdue,Fisher}.
The distribution of the exchange constants is
\[
P(J) = \frac{x\xi_{SP}}{\Delta} \left(\frac{\Delta}{J}\right)^{1-x\xi_{SP}}
\theta\left(\Delta - J\right).
\] 
This distribution
$P(J)$ is such that the typical exchange constant (\ref{T*}),
when $x\xi_{SP}\ll 1$, is much smaller that the average exchange
\begin{equation}
<< J(r) >> =  \Delta \frac{x\xi_{SP}}{1+x\xi_{SP}}
\label{eq:averageJ}
.
\end{equation}
The energy scale $T_*$ is defined by 
\begin{equation}
T_* \simeq {\rm e}^{<<\log[J(r)]>>} = \Delta {\rm e}^{-1/(x\xi_{SP})},
\label{T*}
\end{equation}
and is equal to the typical exchange constant among the spins.
The symbol $<<\dots>>$ denotes a disorder average \cite{Fisher-note}.
The definition of $T_*$ in Eq. (\ref{T*}) in the 1D model is equivalent 
to the one we have introduced in
section \ref{sec:sus} on the basis of a 2D modeling.
Above this temperature, at which the correlation length is of the 
order of $1/x$ (the average distance between the impurities), $\xi(T)$ 
decreases with increasing temperature, 
approaching 
$\xi_{SP}$ when $T\to \Delta$.
As shown on Fig. \ref{fig-corre}, at intermediate
temperatures $T_* < T < T_{SP}$
the AF correlation length is increased by disorder.
Above $T_*$, 
the disordered chain takes advantage of rare fluctuations in which     
many spins get closer than the average distance $1/x$, thus forming 
AF correlated clusters due to the exponential dependence of the 
exchange (\ref{J}) upon the distance. 

The experimental values of the N\'eel temperature (see Table
\ref{table}) are clearly of the order of the average exchange
$<< J >>$, and not of the order of the typical exchange $T_*$.
The same is true for the characteristic exchange $J_{\rm imp}$
in (\ref{eq:fit})
and the (negative) Curie temperature $\theta$ in (\ref{eq:curie}).
The orders of magnitude
of $J_{\rm imp}$ and $\theta$ obtained in section
\ref{sec:sus} are in fact a proof of the relevance of disorder in the
physics of the system.

%
%%%%%%%%%%%%%%%%%%%%%%% FIGURE %%%%%%%%%%%%%%%%%%%%%%%%%%%%%%%%%
%%%%%%%%%%%%%%%%%%%%%%%FIGURE END %%%%%%%%%%%%%%%%%%%%%%%%%%%%%%
%

\subsection{Mean field treatment}
\label{sec:Jperp}
We now include the effect of an interchain coupling $J_{\perp}
\simeq 12$K \cite{neutrons} under the form of 
a mean field treatment based on the 1D model proposed and analyzed in
Refs. \cite{usuno,usdue}. As recalled in section \ref{sec:1D},
the 1D version of our model does develop quasi-long-range
AF fluctuations, which is the reason why AF order will be
easily generated once an interchain coupling $J_{\perp}$
is taken into account.
When the interchain exchange $J_\perp$ is switched on, a true N\'eel long 
range order might develop. The question whether the N\'eel
order is a true long range order or a short range order
will be discussed in more details in the light of the
decimation calculation in section \ref{sec:renor}.
We look in the present section at the mean field response to a
uniform staggered field, and therefore the N\'eel order is
infinite ranged, which is not the case when the interchain
coupling is included beyhond the mean field treatement.

%%%%%%%%%%%%%%%%%%%%%%% FIGURE %%%%%%%%%%%%%%%%%%%%%%%%%%%%%%%%%
\begin{figure}
\centerline{\psfig{file=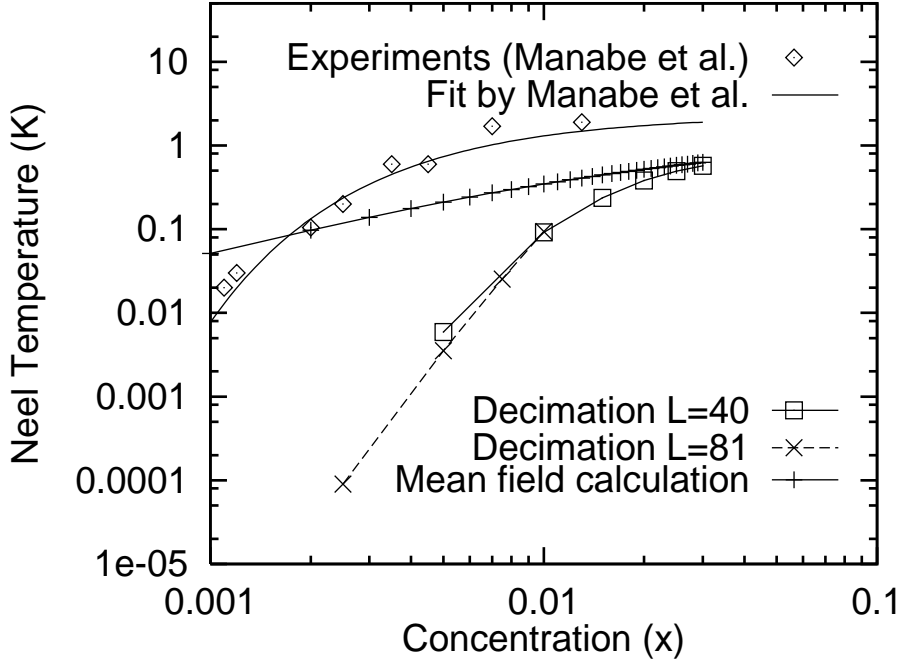,height=8cm}}
\caption{Comparison between the experimental phase
diagram by
Manabe {\it et al.} \protect \cite{Znnew} ($\Diamond$),
our mean-field treatment ($+$) and the decimation results
($L=40$: $\Box$; $L=81$: $\times$).
The fit to the experimental
data is $A \exp{(-B/x)}$, with
$A=2.3 \, \mbox{K}$ and $B=5.7
\times 10^{-3}$.
The mean field results ($+$) have been obtained with $\xi_{SP}=
9 \, \mbox{c}$ and $J_{\perp}=12 \,
\mbox{K}$. The decimation results have been calculated with
sizes $L=40$ ($\Box$) and $L=81$ ($\times$),
and correspond to an existence
probability of $1/2$. As it can be seen on this figure, both
estimations of the N\'eel temperature at low doping are
much larger than $T^* = \Delta \exp{(-1/x \xi_{SP})}$
($\simeq 10^{-47}$K if $x=0.1 \%$). The experimental
results as well as the results from the mean field
treatment and the decimation procedure show no
critical concentration. The mean field and decimation
results have been obtained by using the known parameters
(\ref{eq:parametres}) for CuGeO$_3$ as an input of the calculation,
without adjusting any parameter.
}
\label{fig-Manabe}
\end{figure}
%%%%%%%%%%%%%%%%%%%%%%%FIGURE END %%%%%%%%%%%%%%%%%%%%%%%%%%%%%%
%

To get an estimate of the N\'eel temperature, we 
use the following mean-field like equation for $T_N$: 
\begin{equation}
J_\perp \chi_s(T_N) \xi(T_N) = 1,
\label{eq-TN}
\end{equation}
where $\chi_s(T)$ is the local staggered magnetic susceptibility
(\ref{eq:chis}), and 
$\xi(T)$ the correlation length drawn on Fig. \ref{fig-corre}.
The N\'eel  temperature calculated via
Eq.(\ref{eq-TN}) is plotted on Fig. \ref{fig-Manabe}
as a function of the impurity concentration $x$ for different 
values of $J_\perp$.
It is quite clear from the present calculation
that the N\'eel temperature $T_N$ is much larger that
$T_*$.
This originates from the fact that
large magnetically ordered clusters form already much above $T_*$.
This fact, together with the 
finite polarizability of the singlet dimerized background, 
explains the surprisingly high N\'eel temperatures $T_N$.

We can even obtain a rougher estimate of the N\'eel temperature.
Since  $T_*\ll T_N < T_{SP}$, we can, in a first approximation,
take the correlation
length $\xi(T)$ equal to the spin-Peierls correlation length
$\xi_{SP}$, and further assume that the susceptibility
amounts to a paramagnetic contribution of the free 
magnetic moments:
$\chi_s(T) \simeq x/T.
$
Using Eq.(\ref{eq-TN}) we obtain the N\'eel temperature
\begin{equation}
T_N \simeq J_\perp x\xi_{SP},
\label{TN-rough}
\end{equation}
linear in the impurity concentration. Taking  
$J_\perp\simeq 12 \, \mbox{K}$,
and $\xi_{SP}\simeq 9 \, c$ and $x=0.01$ 
we get $T_N = 1.08 \, \mbox{K}$.
At the same concentration, Martin {\sl et al.} 
measure $T_N\simeq 2\, \mbox{K}$ \cite{Martin}, of the same order
of magnitude as our estimate. 

Both Eqs. (\ref{eq-TN}) and (\ref{TN-rough}) are approximate 
expressions, which might work at
intermediate impurity concentrations, but 
whose validity is doubtful at very low doping $x$.
Anyhow, those results would imply 
that {\it no critical concentration is needed to get a N\'eel phase, since 
a single chain does develop quasi-long range correlations for any 
non zero doping}. Moreover, the N\'eel temperature obtained 
in this mean field treatment is obviously not of the order of
$T_*$, but of the same order of magnitude than the experimental value.
At this point, we judge the mean field treatment to compare
satisfactory to the experimental data, given the fact that
our model is certainly schematic.
These results will be further confirmed by the
decimation analysis in section \ref{sec:renor}.

\section{Low doping physics: decimation in 2D}
\label{sec:renor}
We now present a decimation scheme in 2D which allows to
describe the physics at extremely low doping.
We have already shown in section \ref{sec:Mean-Field} that our
model leads to correct orders of magnitude
of the N\'eel temperature. We do not believe
this mean field theory to be valid at very low doping,
since lowering the doping should
enhance inhomogeneities, and we
then do not expect the physics to be captured by a homogeneous
staggered field involved in the mean-field Stoner criterion
(\ref{eq-TN}). This is why we now consider the full
disordered problem in a fixed dimerized
background.
We take a square lattice of
size $L \times L$, representing the initial Cu
sites of CuGeO$_3$, which, without doping, are paired into a 
spin-Peierls dimerized phase. Some of these
lattice sites will be occupied by non magnetic impurities with a probability
$x$, leaving in their vicinity unpaired spin 1/2 moments. 
The exchange coupling between two such spins at positions 
$(x_1,y_1)$ and $(x_2,y_2)$ is $H_{1-2}=J_{1-2}
{\bf S}_1 \cdot {\bf S}_2$, where we take
\begin{equation}
\label{eq-exchange}
|J_{1-2}| = \Delta \exp{ \left( 
- \sqrt{ \left( \frac{x_2 - x_1}{\xi_x} \right)^2
+        \left( \frac{y_2 - y_1}{\xi_y} \right)^2 } \right) }
,
\end{equation}
$\Delta$ being the spin-Peierls gap $\Delta = 44.7 \, \mbox{K}$, and
$\xi_x \simeq 9 \mbox{c}$ and $\xi_y \simeq 0.1 \,
\xi_x$ the correlation lengths in the $c$ and $b$ directions,
respectively. This model has no adjustable parameter.
The exchange $J_{1-2}$ is negative if
the two sites belong to the same sublattice, and positive 
otherwise, which takes into account the finite response of the 
singlet background to a staggered field. This defines a disordered 
spin model in 2D 
with long range interactions, in the sense that any spin
has an exchange with any other, which however decays
exponentially with their separation.
The decimation procedure we use here is similar
to the one used by Bhatt and Lee  \cite{Bhatt-Lee}
in the study of phosphorus doped silicon. However, 
we decimate strong bonds
without renormalizing the exchanges, which is just sufficient to get 
the order of magnitude of the ordering temperature. 

By construction, our model has bonds which are never frustrated and 
are perfectly compatible with a long range AF order, being ferro on 
the same sublattice and antiferro on different sublattices. Therefore we 
may not get the same low temperature properties found 
by Bhatt and Lee \cite{Bhatt-Lee} in their work on Si:P. 
The reason of the difference is that our 
spins are diluted into a matrix which is easily AF polarizable. A
similar physics does not exist in Si:P.  

\subsection{Decimation in 2D}
In one dimension, a RG transformation of a random AF spin
chain consists in picking-up the strongest exchange
of energy $E$, decimate the two corresponding spins, and
renormalize the exchange by projecting onto the singlet states of the
decimated bonds. 
This leads to broad renormalized 
exchange distributions and to asymptotically
exact low-energy physics \cite{Fisher}.

Similarly to the study on Si:P by Bhatt and Lee \cite{Bhatt-Lee},
we carry out the same procedure in 2D, with the difference
that we do not define a renormalized exchange but just
eliminate the strong bonds. This approximation is 
expected to lead to qualitatively correct results as
far as the estimate of the ordering temperature is
concerned. Considering the 2D disordered problem 
defined by the Hamiltonian (\ref{eq-exchange}),
we decimate the bonds of strength
$E < J \le \Delta$.
Assuming the initial
bonds were colored in white, we color in black these strongly
coupled bonds. As the energy $E$ decreases, more and
more white bonds will be turned into black, until the
black bonds percolate throughout the system at an
energy $E_P$.  We use $E_P$ as an estimate of
the N\'eel temperature $T_N$, even though,
strictly speaking, $E_P$ is an upper bound of
the N\'eel temperature. 
In practice, we use
the following techniques to solve this problem
numerically. (i) The
decimation is carried out by quick sorting the
exchanges.
(ii) A linked list is built at each energy that encodes
the graph of existing exchanges.
(iii) A Hoshen-Kopelman relabeling algorithm \cite{HK} is
used for cluster labeling.
We have tested our program with standard 2D site and bond
percolation, with excellent results.
We now present our results, by first showing in
section \ref{sec:no-self} the failure to find
a percolating cluster in the infinite volume limit.
We next consider in section \ref{sec:phase-diagram}
ordering in a finite coherence volume.

\subsection{Absence of self similar ordering}
\label{sec:no-self}
In this section, we fix the impurity concentration $x$
and look for percolation in an initial lattice of
size $L\times L$, with increasing sizes $L$.
Given the anisotropy of the problem
($\xi_x = 9 = 10 \xi_y$), percolation occurs along the $x$ direction of the
strongest exchanges. It may happen that the left-most
and right-most columns contain no impurity at low
concentration, because of the finite probability
$L x$ to have an impurity present in a given column.
We are thus lead to consider as percolating a cluster
that percolates between the left-most and right-most
columns that contain at least one impurity.

\begin{figure}
\centerline{\psfig{file=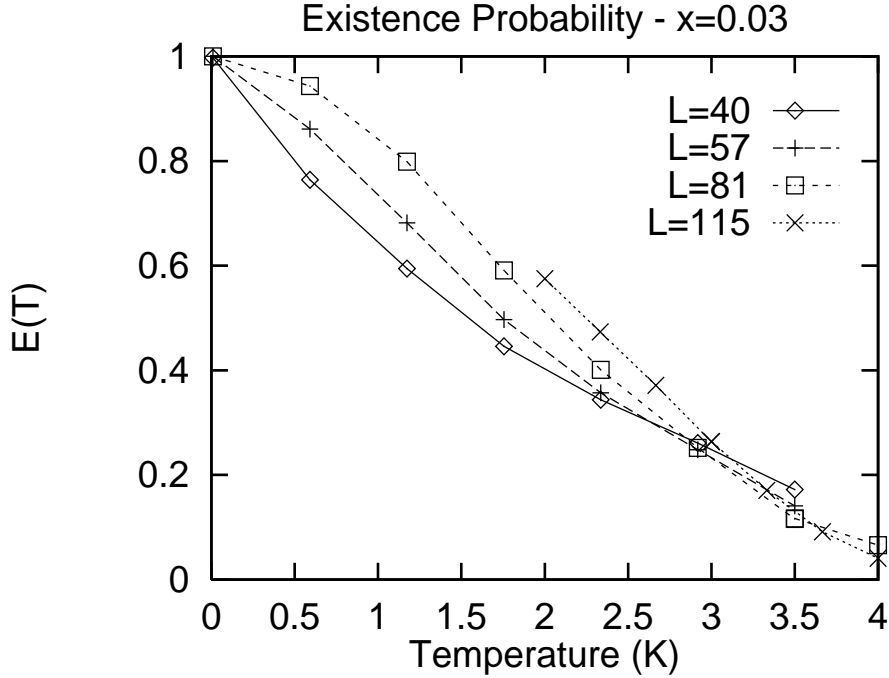,height=8cm}}
\caption{Variations of the existence probability
$E(T)$ versus temperature $T$ at a fixed
impurity concentration $x=0.03$ and for increasing
sizes $L=40$ ($\Diamond$), $L=57$ ($+$), $L=81$ ($\Box$),
$L=115$ ($\times$). As it is visible
on this figure, there is no self-similar fixed point.
}
\label{Existence}
\end{figure}

\begin{figure}
\centerline{\psfig{file=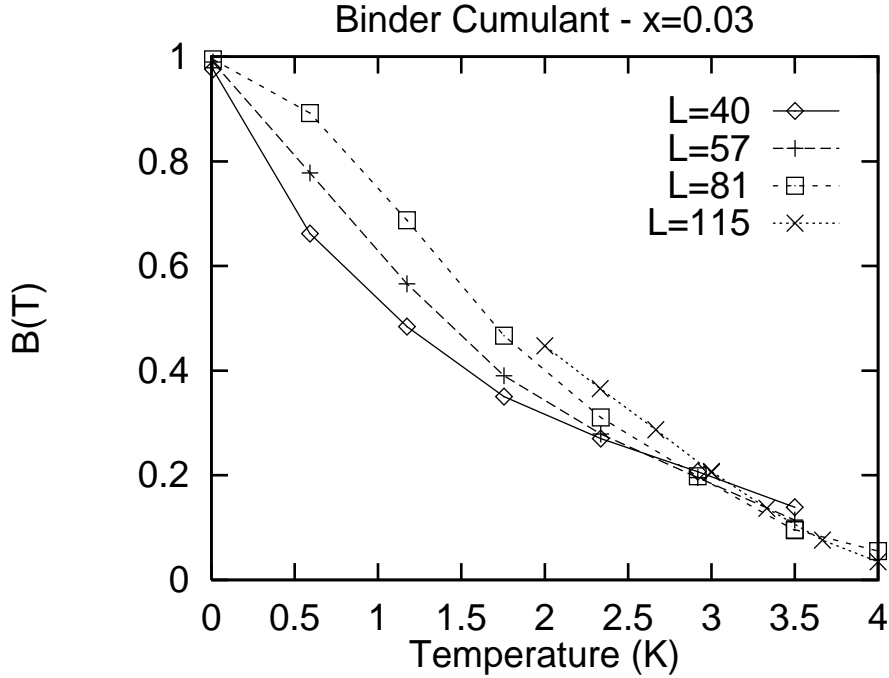,height=8cm}}
\caption{Variations of the Binder cumulant $B(T)$
versus temperature $T$ at a fixed
impurity concentration $x=0.03$ and for increasing
sizes $L=40$ ($\Diamond$), $L=57$ ($+$), $L=81$ ($\Box$),
$L=115$ ($\times$). As it is visible
on this figure, there is no self-similar fixed point, in
agreement with the existence probability result shown on
Fig. \protect \ref{Existence}.
}
\label{Binder}
\end{figure}

In order to determine the percolation probability, we
consider two (equivalent) quantities. First the
existence probability $E_L$, i.e. the probability to find a 
percolating cluster in a system with a finite size $L$.
We also consider the Binder cumulant $B_L = \langle S^2
\rangle_L / \langle S \rangle_L^2$, with $S$ the
number of impurities in a percolating cluster. We analyze 
in this section the finite size scaling behavior of these
two quantities. More precisely, if a true percolation
transition would occur (as it is the case in the 2D
bond and site percolation problem), the crossing points
either of $E_L(T)$ or of $B_L(T)$ for increasing sizes would
accumulate at the transition temperature.
We have plotted on Figs. \ref{Existence} and
\ref{Binder} the existence
probability and Binder cumulant for increasing
sizes at a fixed impurity concentration $x=3 \%$.
As it is visible on these figures, there is no true
percolation transition in the sense that while the crossing point
temperatures are higher and higher as the sizes
increase, the existence probability decreases
to zero. We interpret this result as the lack of  
true long range order in an infinite system. Therefore, we believe that 
AF ordering is of a Griffith type.

%%%%%%%%%%%%%%%%%%%%%%% FIGURE %%%%%%%%%%%%%%%%%%%%%%%%%%%%%%%%%
%%%%%%%%%%%%%%%%%%%%%%%FIGURE END %%%%%%%%%%%%%%%%%%%%%%%%%%%%%%

%%%%%%%%%%%%%%%%%%%%%%% FIGURE %%%%%%%%%%%%%%%%%%%%%%%%%%%%%%%%%
%%%%%%%%%%%%%%%%%%%%%%%FIGURE END %%%%%%%%%%%%%%%%%%%%%%%%%%%%%%

In fact, this finding is not unexpected. At a fixed energy $T$, we can 
interpret the diluted spins as particles 
with a core of area $\sigma$ approximately given by
\begin{equation}
\sigma(T) \simeq \pi \xi_x\xi_y \ln^2\left(\frac{\Delta}{T}\right).
\label{eq:sigma}
\end{equation}
The core is indeed elliptic, with a ratio between the axes 
$a/b = \xi_x/\xi_y$. Hence, the ratio between the 
area occupied by these particles and the surface sample is
\begin{equation}
n_{eff}(T) = x\sigma(T).
\label{eq:eff-density}
\end{equation}   
Ignoring the complication due to the anisotropy, and assuming that
the cores are impenetrable,
a percolating cluster 
can only appear if $T\leq T_P$ where $n_{eff}(T_P)= n_c$, being
$n_c=0.59$ 
the site percolation threshold of a 2D square lattice. 
However, the core of each particle is not impenetrable. For this reason the 
existence probability does not approach one for increasing sizes 
even below $T_P$. Indeed, the existence probability $E(T)$ for 
$L \to \infty$ does not tend to a step function as for 
standard percolation, but to a smooth curve as it is
visible from the finite size study shown on Fig. \ref{Existence}.

\subsection{Low doping phase diagram}
\label{sec:phase-diagram}
The previous analysis suggests that a true N\'eel long range order 
is not established at very low doping, where the singlet 
background is assumed only to provide a coupling between the impurity 
released spins. As we said, for larger doping, the background will be much 
more influenced by the 
impurities, which, besides creating states inside the gap, would also reduce 
the main gap, making a true antiferromagnetic long range order competitive 
with respect to the spin-Peierls phase, as discussed in the Introduction.
 
\begin{figure}
\centerline{\psfig{file=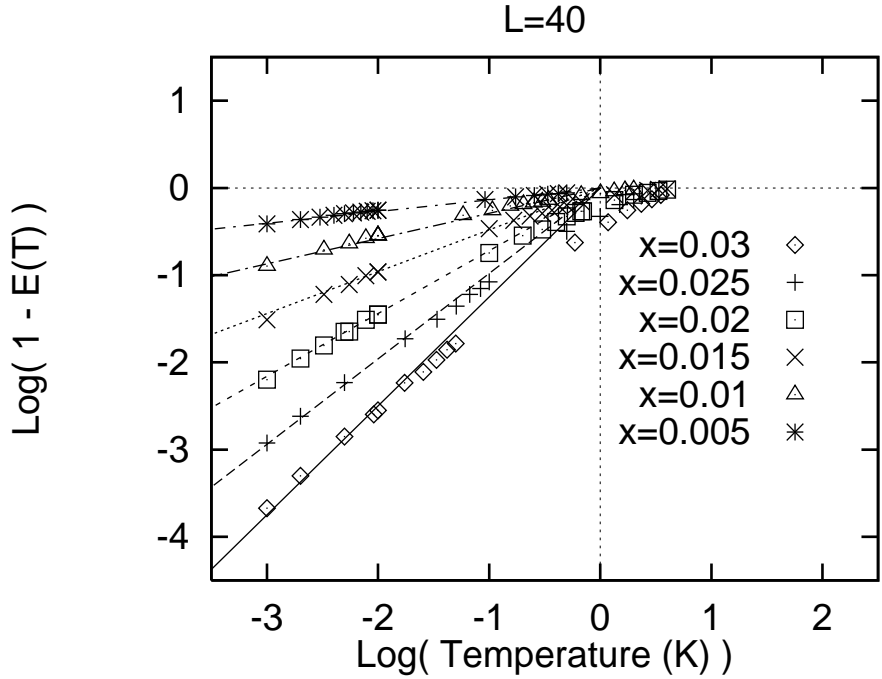,height=8cm}}
\caption{Variations of $1-E(T)$ versus the temperature $T$,
$E(T)$ being the existence probability for percolation
of strongly coupled clusters in a system of size $L=40$.
The different curves correspond to:
$x=0.03$ and $a=1.25$ ($\Diamond$);
$x=0.025$ and $a=0.98$ ($+$);
$x=0.02$ and $a=0.72$ ($\Box$);
$x=0.015$ and $a=0.48$ ($\times$);
$x=0.01$ and $a=0.29$ ($\bigtriangleup$);
$x=0.005$ and $a=0.135$ ($*$).
The CuGeO$_3$ parameters are $\Delta=44.7 \, \mbox{K}$,
$\xi_x=9$ and $\xi_y=0.1 \xi_x$.
}
\label{L40}
\end{figure}

\begin{figure}
\centerline{\psfig{file=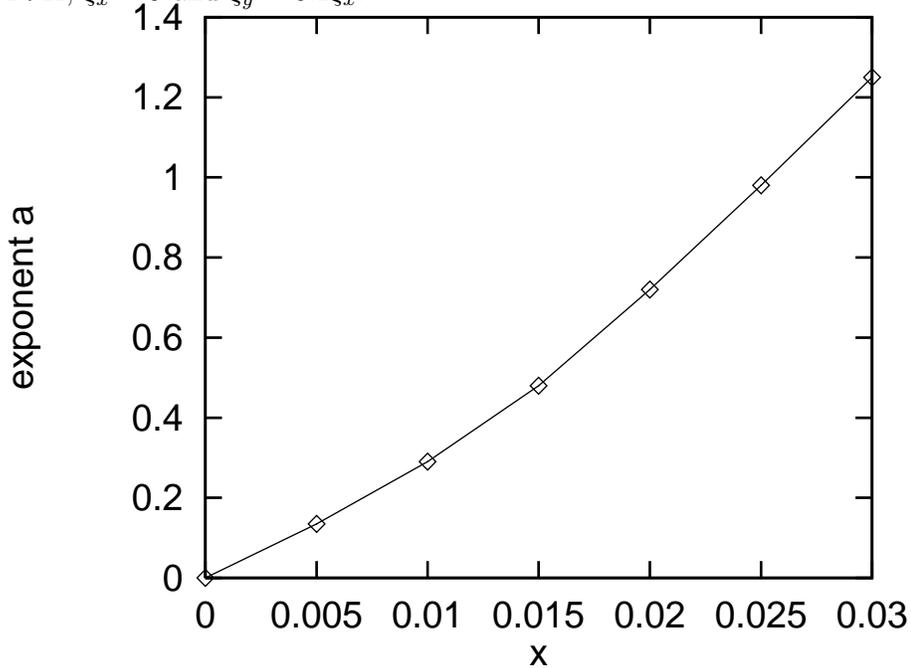,height=8cm}}
\caption{Variations of the exponent $a$ involved in the
fit $1-E(T) = T^a$ shown on Fig. \protect \ref{L40}.
The extrapolation to $x=0$ and $a=0$ has been added and
is consistent with the data at higher concentrations.
}
\label{expo}
\end{figure}

Nonetheless, the neutron beam that probes AF ordering has a finite 
coherence length. Therefore the relevant question, even at low doping, 
is not whether 
a true long range order exists, but rather if clusters of size larger 
than that coherence length appear, and with which probability. 
For this reason, we want now to discuss AF ordering in a finite coherence 
volume, which, as we said, is set either by the size of the sample or 
by the finite coherence length of the experimental probe, e.g. the neutron 
beam. 
Therefore we vary the impurity concentration $x$ at fixed
size $L$. The N\'eel temperature is estimated by imposing
a fixed value of the existence probability $E_L(T)=1/2$. 
We are thus forced to analyze the low temperature behavior of the
existence probability $E_L(T)$. This is shown on
Fig. \ref{L40} for $L=40$ and impurity concentrations
between $0.5 \%$ and $3 \%$. With these parameters, the
low temperature behavior of the existence probability
is well fitted by the power-law $1 - E_L(T) = T^{a(x)}$.
Fig. \ref{expo} shows the variations of the
exponent $a(x)$ versus $x$.
In this figure, we have added
the point $(x=0, a=0)$, and, as it is visible,
this extrapolation is compatible with the large $x$ data, hence 
showing that
zero temperature AF ordering occurs for an
infinitesimal doping in a finite size. Imposing a fixed existence
probability $E_L(T_N)=1/2$, we obtain 
$\ln{T_N} \propto -1/x$, a form of the N\'eel temperature
proposed from the low doping measurements in Ref.
\cite{Znnew}. The calculated N\'eel
temperature is shown on Fig. \ref{fig-Manabe}, together
with the experimental points by Manabe {\it et al.}.
Compared to these experiments, we 
obtain a smaller value of the N\'eel temperature, still
much above $T_*$, in agreement with \cite{Znnew}.

%%%%%%%%%%%%%%%%%%%%%%% FIGURE %%%%%%%%%%%%%%%%%%%%%%%%%%%%%%%%%
%%%%%%%%%%%%%%%%%%%%%%%FIGURE END %%%%%%%%%%%%%%%%%%%%%%%%%%%%%%

%%%%%%%%%%%%%%%%%%%%%%% FIGURE %%%%%%%%%%%%%%%%%%%%%%%%%%%%%%%%%
%%%%%%%%%%%%%%%%%%%%%%%FIGURE END %%%%%%%%%%%%%%%%%%%%%%%%%%%%%%

\begin{figure}
\centerline{\psfig{file=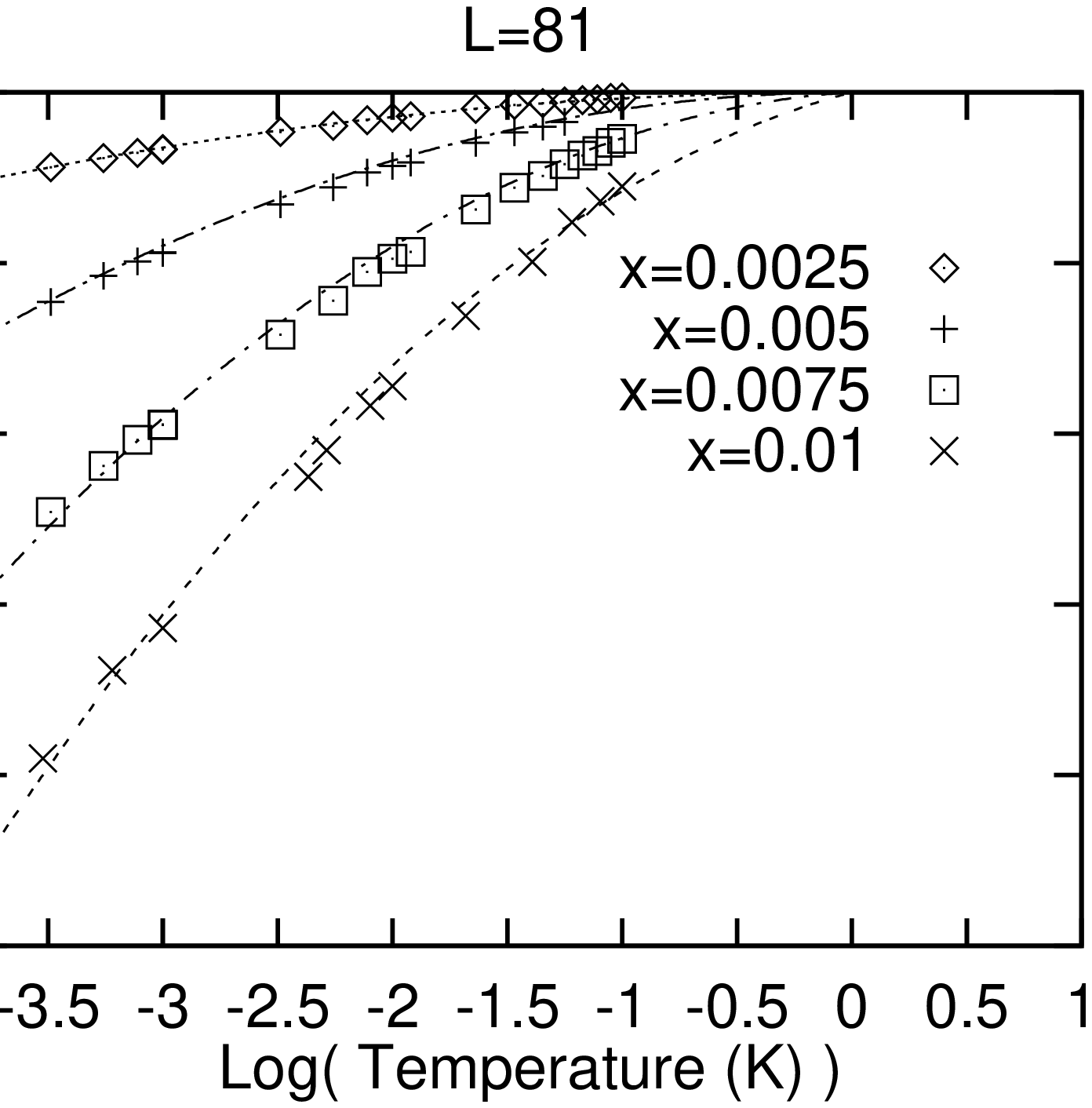,height=8cm}}
\caption{Variations of $1-E(T)$ versus the temperature $T$,
$E(T)$ being the existence probability for percolation
of strongly coupled clusters in a system of size $L=81$.
The fits are of the form $\log{(1-E(T))}  = f(\log{(T)})$, with
$x=0.01$: $f(x)=-0.11 x^2 + 0.18 x$ ($\times$);
$x=0.0075$: $f(x)=-0.092 x^2 + 0.042 x$ ($\Box$);
$x=0.005$: $f(x) = -0.05 x^2$ ($+$);
$x=0.0025$: $f(x)=-0.018 x^2$ ($\Diamond$).
The CuGeO$_3$ parameters are $\Delta=44.7 \, \mbox{K}$,
$\xi_x=9$ and $\xi_y=0.1 \xi_x$.
}
\label{L81}
\end{figure}

In order to analyze the effect of increasing the
coherence length, we now consider a system with a 
larger size $L=81$. The low temperature behavior of the
existence probability is shown on Fig. \ref{L81}, with
doping concentrations from $0.25 \%$ to $1 \%$, where
the inclusion of a quadratic term was necessary in
the fit.
The N\'eel temperature obtained by imposing 
$E_L(T_N) = 1/2$ is shown on Fig. \ref{fig-Manabe},
together with the $L=40$ result and the experimental
data. As it is visible on this figure, in spite of the
different behavior of the low temperature existence
probability $E_L(T)$ when the system size was
increased from $L=40$ to
$L=81$, the orders of magnitude of the N\'eel temperature
$T_N(x)$ remain similar.

%%%%%%%%%%%%%%%%%%%%%%% FIGURE %%%%%%%%%%%%%%%%%%%%%%%%%%%%%%%%%
%%%%%%%%%%%%%%%%%%%%%%%FIGURE END %%%%%%%%%%%%%%%%%%%%%%%%%%%%%%

The above analysis suggests that at low doping the 
model does not have a true long range order in the 
thermodynamic limit, but the probability to find 
an ordered cluster of any given size $L$ is finite. This implies that 
the system will appear as if it were ordered to an external probe 
with a finite coherence length. This scenario gives a natural explanation 
of the fact that the measured N\'eel temperatures are always much 
larger than the {\sl typical energy scale} $T_*$ of the doped 
CuGeO$_3$, and of the evidence that the intensity of the AF peaks 
observed in neutron scattering is not saturated even at 
$1.4 \,\mbox{K}$ \cite{Hase} for a Zn doped sample with $x=0.034$.   
  
To obtain the dependence of the susceptibility upon
temperature, we would need to improve our decimation scheme
by renormalizing the exchanges, 
similarly to  Ref. \cite{Bhatt-Lee}, and inserting a weak spin anisotropy, 
to break the spin symmetry. 

Let us summarize our decimation results by comparing them
to the experiments by Manabe {\it et al.} \cite{Znnew}, and
to $T_*$:
\begin{eqnarray}
\mbox{Fit by Manabe {\it et al.}:} && \ln{T_N} \sim 5.7 
\times 10^{-3}/x\\
\mbox{Decimation ($L=40,81$):} && \ln{T_N} \sim 10^{-3}/x \\
\mbox{Behavior of $T_*$:} && \ln{T_*} \sim 10^{-1}/x
,
\end{eqnarray}
which illustrates both the fact that we can reproduce
sucessfully the absence of critical concentration in
our model, and the crucial role of disorder in
enhancing the N\'eel temperature.

\section{Conclusion and discussion}
\label{sec:conclusion}
Let us summarize the proposal we have  
presented for the onset of the N\'eel long range order in doped CuGeO$_3$.
We first want to stress that, in our opinion, 
the real difficulty in the subject is
not the coexistence of dimerization and antiferromagnetism,  
since the two order parameters are not incompatible. 
Instead, what is puzzling is 
the appearance of antiferromagnetism at such low impurity 
concentration as $1\times 10^{-3}$, with a reasonably high transition 
temperature. We have shown that this behavior might depend on the 
peculiar properties of the disorder in this system, as well as in other 
spin gaped  systems.

We believe that the most important effect of disorder, at low doping, 
is to induce randomly distributed 
localized free moments. These moments are  
weakly coupled by an exchange constant 
exponentially decaying upon the distance, see 
Eq.(\ref{J}), caused by the virtual polarization of the singlet background. 
We have shown that even a single chain does develop quasi long range 
N\'eel order, {\it i.e.} a power law decaying 
staggered correlation function, 
which, we believe, does imply a low temperature transition to 
an AF phase as soon as 
the interchain exchange is taken into account. 
Moreover, we argued that the exponential dependence on the 
distance of the exchange between those randomly distributed moments implies 
that rare events in which the 
impurity released spins are closer than the average distance
(and therefore get strongly coupled), dominate the physics
at low doping, in a Griffith-like scenario. This picture was
fully confirmed by an investigation of the 2D disordered
problem using a decimation method.
The formation of these Griffith-like clusters, 
together with the finite staggered susceptibility of the singlet background, 
may explain why ordered domains larger that the neutron coherence length 
do appear at temperatures much larger than the temperature $T_*$ 
(the exchange between two moments at a distance equal to the 
average one). 

Some consequences can be drawn from this scenario:
\begin{itemize}

\item  A N\'eel transition is expected for any 
arbitrary small impurity concentration.

\item At relatively large doping concentrations, the 
mean field picture is expected to hold since the inhomogeneities
of the AF order are weak. In this concentration
range, the N\'eel temperature behaves like
$T_N \simeq J_b \xi_{SP} x$, where $J_b$ is the exchange constant 
in the $b$-direction, and $\xi_{SP}$ the correlation length in the pure 
CuGeO$_3$. 

\item For low impurity concentrations, the N\'eel state
becomes highly inhomogeneous and the decimation method is
more suited than the mean field theory. Instead of a 
single transition temperature we find a quite broad distribution of $T_N$'s. 
We have indications that a true long range order is never 
established, so that the low temperature 
phase might be rather interpreted as a
Griffith-like antiferromagnet.
We have shown that, for 
small coherence volumes, the N\'eel temperature behaves like
$\ln{T_N} \sim -1/x$. By increasing the coherence volume, this
behavior does not rigorously hold. However, as it is visible on Fig.
\ref{fig-Manabe}, the behavior of the N\'eel temperature
for $L=81$ can hardly be discriminated from the $L=40$ calculation,
and a fit of our results
to a $\ln{T_N} \sim -1/x$ behavior
would be equally satisfactory.

\item A very long correlation length in the c-axis direction, 
extending well above $\xi_{SP}$, may be observed
in a wide temperature range. 
This can be measured in NMR experiments at very low doping.
For instance, the NMR linewidth in the ladder compound 
Sr(Cu$_{1-x}$Zn$_x$)$_2$O$_3$ at 
a Zn concentration $x=0.25\%$, at which no N\'eel transition has been 
observed down to the lowest accessible temperature, is   
two orders of magnitude 
larger than the correlation length of the undoped compound \cite{NMR},
which indicates that a similar behavior might well occur
in the doped CuGeO$_3$.
\end{itemize}

We now compare our results with other theoretical proposals.  
We first compare our work to the one by Mostovoy {\it et al.}
\cite{Mostovoy}, and next to the work by Fukuyama
{\it et al.} \cite{Fuku}, and Yoshioka and Suzumura \cite{Suzumura}.
The technicalities involved in the discussion of these last
two works are left in the Appendices \ref{ap:SCHA} and
\ref{ap:solitons}, and we focus here in the concluding section
on more qualitative aspects.

Mostovoy {\it et al.} \cite{Mostovoy}
have recently investigated the effects of interchain coupling at a
mean field level in an antiferromagnetic Heisenberg 
chain with random dimerization, modeled by a $\delta$-correlated in space 
gaussian-distribution peaked around a finite average value. In order to make
contact with their work, we make the following comments:
\begin{itemize}
\item[(i)] The microscopic modeling of doping in a spin-Peierls
system is quite different in our model and in Ref. \cite{Mostovoy}.
As we showed in Ref. \cite{usuno}, our model is more related to 
a dimerization with a random telegraph noise of 
zero average distribution, with the low temperature
physics dominated by the 
localized spins around the domain walls.
\item[(ii)] The physics of the 1D limit of both models is different.
This can be seen by considering the RG flow
\cite{Dasgupta-Ma,Fisher}. The 1D limit of the
model by Mostovoy {\it et al.} has been solved by Hyman
{\it et al.} \cite{Hyman}, and flows to a fixed point
with power-law Griffith singularities. By contrast, the 1D
limit of our model flows to the random singlet fixed point
\cite{Fisher}
with logarithmic singularities \cite{usuno,usdue}. 
\item[(iii)] More importantly, the random dimerized Heisenberg chain,
which is the starting point of the description in Ref.
\cite{Mostovoy}, does {\it not} develop quasi long range
AF fluctuations, as it is the case in our proposal.
This allows to understand why a finite critical concentration
of impurities is required in Ref. \cite{Mostovoy} to generate
an AF-dimerized phase. As we show in the present article,
this is not the case in our model, and it seems that this
is not the case in experiments either \cite{Znnew}.
\end{itemize}

A proposal similar to the one by
Mostovoy {\it et al.} was previously put forward 
by Fukuyama, Tanimoto and Saito (FTS) \cite{Fuku}, 
who also worked out a model showing a coexistence of
antiferromagnetism and  dimerization, as well as the 
onset of antiferromagnetism for arbitrary small doping. As we discussed, 
we have started our analysis by assuming that the main effect of the 
impurities is the release of  
free spins out of the singlet background, which is in agreement with 
the magnetic susceptibility data on the doped compounds, showing an almost 
Curie like component at low temperature which seems 
to scale linearly with doping (see for instance 
Ref. \cite{Grenier} and \cite{Znnew}). 
Instead, FTS assume, as Mostovoy {\it et al.}, that the main effect of 
disorder is the local reduction of the dimerization 
$\delta_{imp}<\delta$ close to each impurity, hence of 
the spin gap.  For instance, they can fit the experimental reduction of 
the scattering intensity from the dimerized lattice in the 
$0.7\%$ Si doped compound, by taking 
$\delta_{imp}\sim 0.2 \delta$. This leads to a rough estimate of the 
local reduction of the spin gap $\Delta_{imp}\sim (0.2)^{2/3}\Delta 
\simeq 0.9 \Delta$. Obviously, for smaller impurity concentrations, one 
should expect an even smaller reduction of the spin gap. 
However, to explain within this model the almost 
Curie like behavior of the low temperature susceptibility at low doping,   
one should assume a much stronger suppression of the spin gap down to 
$\sim 10 \mbox{mK}$ close 
to each impurity, not just a small reduction as the fit to the 
neutron scattering data would imply. This is one reason why we 
believe that our approach is more relevant. The other is 
that we do not believe that a local reduction of the dimerization 
might explain the appearance of antiferromagnetism at such low concentration 
as $0.1\%$ of Zn doping.    

In fact, a more realistic modeling of disorder
in CuGeO$_3$ should include 
both disorder effects, the local reduction of dimerization as well as 
the appearance of free spins. Essentially, the main difference 
between our approach and 
the FTS one is in the different emphasis placed on the two effects. 
We believe that 
the main role at low doping in establishing the N\'eel phase is played 
by the free moments induced by disorder, while FTS seem to favor the other 
mechanism of a local reduction of the dimerization.    
 
Apart from this comment on the grounds of the model proposed in
Ref. \cite{Fuku}, we also have some more technical criticisms.     
To analyze their model, FTS use the so-called 
Self Consistent Harmonic Approximation (SCHA) applied to the bosonized 
version of a dimerized Heisenberg chain. This approximation is a 
mean field theory, or saddle point approximation, plus random phase 
fluctuations. By this technique, they 
show the existence of a mean field solution having both 
a finite dimerization parameter and a finite N\'eel order
parameter. Essentially, close 
to the impurity, where the dimerization parameter is assumed to be reduced,
a staggered magnetic moment in the $z$-direction arises. 
We believe that this approach is not fully consistent, as we explain  
in detail in Appendix A. 

An alternative variational approach has recently been developed by 
Yoshioka and Suzumura \cite{Suzumura} for the Zn doping case. 
They have rigorously shown that in bosonization the spin-lattice coupling 
does change sign crossing a non magnetic impurity, as it was proposed in 
Ref. \cite{usuno}. With this disorder modeling they have proved the 
existence of a variational action which does again describe a state where 
the dimerization gets reduced around an impurity and at the same time 
a static staggered moment appears in the $z$-direction.
We believe that this 
variational approach gives a reasonable description of the 
N\'eel ordered phase, very close to what we propose, but does
not allow a description of how the AF ordering occurs,
which we did  in the present work.  
We postpone to Appendix \ref{ap:solitons} a detailed
discussion of this point, and we 
end-up with two simple comments on these semiclassical analyzes. 

The dimerized phase breaks translational invariance like a N\'eel phase does, 
which, in addition, also breaks spin symmetry. If one analyzes the role 
of defects in a dimerized state within a scheme which breaks 
SU(2) symmetry (what is implicitly done in those semiclassical approaches,
as shown in Appendix \ref{ap:solitons}), 
it is not a surprise that a finite staggered magnetization comes out of the 
calculation (without spin anisotropy or an applied staggered magnetic field),
since the dimerized state already breaks translational 
symmetry.

Finally, one can work out a model in the spirit of the FTS
scenario. This can be done
by considering a dimerization distributed as a telegraph noise,
and taking a value $\delta$ in the dimerized background (large segments),
and $0 < \delta_{imp} < \delta$ close to the impurities (short segments).
The resulting disordered model can be exactly solved in the XX limit
along the lines in Ref. \cite{Monthus}. This would result in a model
with power-law Griffith singularities, equivalent to the model
by Mostovoy {\it et al.} \cite{Mostovoy}, and inequivalent to our
model where the Dyson singularities are power-law. This model
would not flow to the random singlet fixed point \cite{Fisher},
would have much weaker AF fluctuations than our model, and,
as in the case of the model by Mostovoy {\it et al.},
a critical concentration would be expected for the appearance
of AF order.

\section{Acknowledgments}
We acknowledge useful discussions with J.C. Angl\`es d'Auriac,
C. Berthier, D. Fisher, T. Giamarchi,
B. Grenier, M. Horvatic, J. C. Lasjaunias,
J. Lorenzo, P. Monceau,
A. Nersesyan, Y. Okabe,
C. Paulsen, J.P. Pouget, G. Remenyi and M. Saint-Paul
at various stages of this project.
The authors also thank K. Uchinokura for sending them
their susceptibility data.
This work has beeen partly supported by INFM,
Research Project HTSC.
The decimation calculations have been performed on the CRAY
T3E supercomputer of the Centre de Calcul Vectoriel
Grenoblois of the Commissariat \`a l'Energie Atomique.

\appendix

\section{Self Consistent Harmonic Approximation}
\label{ap:SCHA}
In this Appendix, we discuss the 
Self Consistent Harmonic Approximation (SCHA) 
approach used by Fukuyama, Tanimoto and Saito (FTS) in Ref. \cite{Fuku}. 
We do not want to enter in the details of the transformation which allows 
to map the spin-Peierls problem onto a bosonic one, since they can be  
found in Ref. \cite{Fuku}, and references 
therein. We assume that this transformation is valid, and, following 
\cite{Fuku}, that the  
resulting bosonic Hamiltonian is 
\begin{equation}
\hat{H} = \frac{1}{2} \int dx \left[ 
\Pi(x)^2 + \left(\partial \phi(x)\right)^2 
-2gu(x)\sin\left(\sqrt{4\pi K}\phi(x)\right)
+ \kappa u^2(x) \right],
\label{Hamfull}
\end{equation}
where $\phi(x)$ and $\Pi(x)$ are bosonic conjugate fields. 
The classical variable $u(x)$ represents the dimerization lattice distortion, 
and the $\sin(\dots)$ is the spin dimerization order parameter. 
Notice that the $\cos(\dots)$ would represent the $z$-component of the 
staggered magnetization. 
We then write $\phi(x) = \phi_c(x) + \phi(x)$, where $\phi_c(x)$ is a 
classical variable, and $\phi(x)$ is the quantum fluctuation part, which 
is assumed to have a zero average value. We want to find the minimum of the 
total energy assuming a bosonic state $|0\rangle$ 
which is the ground state of the following Hamiltonian
\begin{equation}
\hat{H}_{sc} = \frac{1}{2}\int dx \left[
\Pi(x)^2 + \left(\partial \phi(x)\right)^2 + m(x)\phi^2(x)
\right]
,
\label{Hamsc}
\end{equation}
with $m(x)>0$.
Therefore we have to find the appropriate values of $\phi_c(x)$, 
$u(x)$ and $m(x)$ which minimize the average value of $\hat{H}$ 
over $|0\rangle$. 
This is the so-called Self Consistent Harmonic Approximation (SCHA), which 
is a well defined variational technique. 
Since (\ref{Hamsc}) is quadratic, then 
$
\langle 0| \phi(x) |0\rangle = 0,
$
and 
\[
\langle 0| \sin \left(
\sqrt{4\pi K}\left(\phi_c(x)+\phi(x)\right)\right) |0\rangle = 
D(x)\sin\left(\sqrt{4\pi K}\phi_c(x)\right),
\]
where 
$
D(x) = {\rm e}^{-2\pi K \langle \phi(x)^2 \rangle}.
$
If we define 
$
T(x) \equiv \langle \Pi^2(x) + 
\left(\partial \phi(x)\right)^2  \rangle, 
$
the self-consistent equations are 
\begin{eqnarray}
&& -\partial^2\phi_c(x) - g \sqrt{4 \pi K} u(x) D(x) 
\cos{\left( \sqrt{4\pi K}\phi_c(x) \right)}=0; 
\label{eq-a}\\
&& \kappa u(x) = g D(x)\sin{\left(\sqrt{4\pi K}\phi_c(x)\right)};
\label{eq-b}\\
&& \int dy \frac{\delta T(y)}{\delta m(x)} 
+ 4 \pi K g u(y)
\sin{\left(\sqrt{4\pi K}\phi_c(y)\right)} D(y) 
\frac{\delta\langle \phi(y)^2 \rangle }{\delta m(x)}=0, \label{eq-c} 
\end{eqnarray}
where we used 
\[
\frac{\delta D(y)}{\delta m(x)} = -2 \pi K D(y)
\frac{\delta \langle \phi(y)^2 \rangle }{\delta m(x)}.
\]
We start by solving (\ref{eq-c}). We write the bosonic fields as 
\begin{eqnarray}
\phi(x) = \sum_\alpha \frac{1}{\sqrt{2\epsilon_\alpha}}
\varphi_\alpha(x) \left(b_\alpha + b^\dagger_{-\alpha}\right),
\label{phi} \\
\Pi(x) = i\sum_\alpha \sqrt{\frac{\epsilon_\alpha}{2}}
\varphi^*_\alpha(x) \left(b^\dagger_\alpha - b_{-\alpha}\right),
\label{Pi} 
\end{eqnarray}
where $\varphi_{-\alpha}(x)=\varphi^*_\alpha(x)$, 
$\epsilon_\alpha = \epsilon_{-\alpha}$, 
\[
\sum_\alpha \varphi^*_\alpha(x) \varphi_\alpha (y) = \delta(x-y),
\]
and $[b_\alpha,b^\dagger_\beta]=\delta_{\alpha,\beta}$, all other 
commutators being zero. We want the Hamiltonian (\ref{Hamsc})
to be diagonal in terms of these 
creation and annihilation operators: 
\[
\hat{H}_{sc} = \sum_\alpha \epsilon_\alpha b^\dagger_\alpha b_\alpha.
\]
A straightforward calculation gives
\begin{equation}
-\partial^2 \varphi_\alpha (x) + m(x)\varphi_\alpha (x)= 
\epsilon^2_\alpha \varphi_\alpha (x), 
\label{eq-phi}
\end{equation}
which is the eigenvalue equation for $\varphi_\alpha$. On the bosonic vacuum 
of the operators $b_\alpha$, we find that 
\[
\int dy \, T(y) = \frac{1}{2} \sum_\alpha \int dy \left[
\epsilon_\alpha \, \varphi^*_\alpha(y) \varphi_\alpha (y) 
+ \frac{1}{\epsilon_\alpha} \partial \varphi^*_\alpha (y) 
\partial \varphi_\alpha (y)\right], 
\]
which, through Eq.(\ref{eq-phi}), becomes
\begin{eqnarray*}
\int dy T(y)&=& \sum_\alpha \frac{1}{2 \epsilon_\alpha} \int dy    
\left(2 \epsilon^2_\alpha - m(y)\right) 
\varphi^*_\alpha (y) \varphi_\alpha (y)\\
&=& \sum_\alpha \left[ \epsilon_\alpha - 
\int dy \frac{m(y)}{2 \epsilon_\alpha}\varphi^*_\alpha (y)
\varphi_\alpha (y) \right]\\
&=& \sum_\alpha \epsilon_\alpha - \int dy m(y) \langle \phi^2(y) \rangle,
\end{eqnarray*}
where the last identity follows from (\ref{phi}). 
Now, suppose that $m(x)\to m(x) + \delta m(x)$. It is easy to 
get the first order correction to the eigenvalues through (\ref{eq-phi}):
\begin{eqnarray*}
\sum_\alpha \delta \epsilon_\alpha &=& \sum_\alpha 
\frac{1}{2\epsilon_\alpha}
\int dy\, \varphi^*_\alpha(y)\, \delta m(y)\, \varphi_\alpha (y) \\
&=& \int dy\, \delta m(y) \langle \phi^2(y) \rangle.
\end{eqnarray*}
Through the above equations we find that 
\[
\int dy\, \frac{\delta T(y)}{\delta m(x)} = 
\langle \phi^2(x) \rangle  - \left[ \langle \phi^2(x) \rangle
+ \int dy \, m(y) \frac{\delta \langle \phi^2(y) \rangle}{\delta m(x)}\right],
\]
which, inserted in Eq.(\ref{eq-c}) gives
\[
\int dy \left[ - m(y) + 4 \pi K gu(y)
\sin\left(\sqrt{4\pi K} \phi_c(y)\right) D(y)\right]
\frac{\delta \langle \phi^2(y) \rangle}{\delta m(x)} = 0.
\]
This equation is solved by imposing 
\begin{equation}
m(x) = 4 \pi K g u(x)
\sin\left(\sqrt{4\pi K}\phi_c(x)\right) 
{\rm e}^{-2\pi K \langle \phi(x)^2 \rangle},
\label{eq-mass}
\end{equation}
which gives the self-consistent equation for the mass. 

As a simple application, let us consider the homogeneous case
$m(x) = m$, $u(x)=u$, and $\phi_c(x) =\phi_c$ for any $x$.
From Eq. (\ref{eq-a}) we get 
\begin{equation}
\sqrt{4\pi K}\phi_c(x) = \pi (n+\frac{1}{2}), 
\label{SU2}
\end{equation}
which choice is in fact the only compatible with SU(2) symmetry. 
We choose $n=1$. On the other hand Eq.(\ref{eq-phi}) is solved by 
plane waves 
\[
\varphi_q(x) = \frac{1}{\sqrt{L}}{\rm e}^{-iqx},
\]
with energy $\epsilon_q = \sqrt{q^2 + m}$. Therefore
\[
\langle \phi(x)^2 \rangle = \frac{1}{2L}\sum_q \frac{1}{\sqrt{q^2 + m}}
\simeq \frac{1}{4\pi}\log \left(\frac{m_0}{m}\right),
\]
where $m_0$ is a high-energy cut-off, and (\ref{eq-mass}) becomes
\[
m = 4 \pi K g u  
\left(\frac{m}{m_0}\right)^{K/2}.
\]
The solution is 
$
m \sim (gu)^{2/(2-K)}.
$
Since the approach is valid only if, for a small coupling $g$, a small mass 
$m$ is obtained, the above equation implies that a mass is generated 
if $K<2$. This is the famous result for the sine-Gordon models. If we 
consider this model as describing a spin-Peierls transition, we 
know that $K=1$ corresponds to the XX-limit and $K=1/2$ to the 
SU(2) point. In the latter case, 
\[
m = 2 \pi g u  
\left(\frac{m}{m_0}\right)^{1/4}.
\]
Eq.(\ref{eq-b}) is solved by 
\[
u = \frac{g}{\kappa}\left(\frac{m}{m_0}\right)^{1/4},
\]
finally leading to 
\begin{equation}
m = \frac{4 \pi^2 g^4}{\kappa^2 m_0}.
\label{m-unif}
\end{equation}

Now let us consider the inhomogeneous case. For instance, following 
FTS \cite{Fuku}, we assume that $u(0)=u(l)=u_i$. Through (\ref{eq-mass})   
the self consistent equations can be written as    
\begin{eqnarray}
&& -\partial^2\phi_c(x) - g \sqrt{4 \pi K}
u(x) D(x) \cos{\left(\sqrt{4\pi K}\phi_c(x)\right)}=0; 
\label{eq-a1}\\
&& \kappa u(x) = gD(x)\sin{
\left(\sqrt{4\pi K}\phi_c(x)\right)}; \label{eq-b1}\\
&& m(x) = 4 \pi K gu(x)
\sin{\left(\sqrt{4\pi K}\phi_c(x)\right)} D(x).
\label{eq-c1}
\end{eqnarray} 
Solving for $u(x)$, we get 
\[
u^2(x) = \frac{1}{4 \pi K \kappa} m(x),
\]
and the coupled equations for $m(x)$ and $\phi_c(x)$ read
\begin{equation}
m(x) = \frac{4 \pi K g^2}{\kappa} D^2(x) 
\sin^2\left(\sqrt{4\pi K}\phi_c(x)\right),
\label{m-non-unif}
\end{equation}
\begin{equation}
-\partial^2\phi_c(x) - \frac{g^2 \sqrt{4 \pi K}
D^2(x)}{2\kappa}
\sin{\left(2\sqrt{4\pi K}\phi_c(x)\right)}=0 .  
\label{equno}
\end{equation} 
We do not know whether these equations have an unique solution. 
FTS implicitly assume that $D(x)$ as well as 
$m(x)$ are independent of $x$. 
Then Eq.(\ref{eq-b1}) at $x=0,l$ fixes the value of $\phi_c(0)$ 
and $\phi_c(l)$, which are used as boundary conditions to solve 
Eq.(\ref{equno}). The solution is a Jacobi elliptic function. 
Between $x=0$ and $x=l$, the solution approaches the homogeneous one 
(\ref{SU2}), but around these points it is different. 
As a result, $\sin^2\left(2\sqrt{4\pi K}\phi_c(x)\right)$ 
differs from one close to $x=0$ and $x=l$, leading to a 
finite value of $\cos^2\left(2\sqrt{4\pi K}\phi_c(x)\right)$, hence to 
a finite staggered magnetization. This result would imply that 
disorder in the lattice distortion leads automatically to a long 
range N\'eel phase, but necessarily with a staggered magnetization along 
$z$. In fact, the staggered magnetization perpendicular to $z$ 
is expressed in terms of the conjugate momentum $\Pi(x)$, which 
is uncertain if $\phi_c(x)$ gets an average value. This result is puzzling, 
since the starting Hamiltonian is spin isotropic.     
Moreover, the assumptions $D(x)$ and $m(x)$ $x$-independent have to 
be {\it aposteriori} verified. FTS do not make this check, but 
clearly, by inspection of Eq.(\ref{m-non-unif}), one realizes that 
the assumption is not correct. 
Therefore the calculations of 
Ref. \cite{Fuku} are not self-consistent. 

Alternatively, one can look for a 
solution which does not break SU(2) symmetry. In this case we have to 
assume a constant $\phi_c(x)=\pi/2$, 
compatible with (\ref{SU2}). Therefore one needs to solve
\begin{eqnarray}
&& \kappa u(x) = gD(x); \label{eq-b2}\\
&& m(x) = 2 \pi g u(x) D(x), \label{eq-c2}
\end{eqnarray}
or, equivalently, 
\[
m(x) = \frac{2\pi}{\kappa} g^2 D^2(x) = 2\pi \kappa u^2(x),
\]
with the boundary conditions $m(0)=m(l)=2\pi \kappa u^2_i$. 
If this equation were solvable, we would get an alternative 
solution which keeps the SU(2) symmetry. Further work to establish which 
is the best self-consistent solution is needed, but we can safely state that 
the assumption of homogeneous quantum fluctuations made in 
Ref. \cite{Fuku} is not self-consistent.

\section{Some rigorous results in the presence of solitons}
\label{ap:solitons}
In order to better clarify the limits of the SCHA, we consider a simple case 
which is related to the work by Yoshioka and Suzumura \cite{Suzumura}. 
These authors have proved by bosonization that the coupling 
to the lattice changes sign crossing a non magnetic 
ions. A similar conclusion was reached in Ref. \cite{usuno}, by 
means of a simple mapping to a squeezed chain where the non magnetic sites 
are eliminated. This mapping is explained in  
Fig.\ref{mapping}. The squeezed chain is obtained by eliminating the 
non magnetic site similarly to what is done in the $U\to\infty$ 
Hubbard model. The difference here is that the {\it hole} does not move. 
Next, the weak link across the impurity is approximated as the weak bond 
of the dimerized pattern, {\it i.e.}
$J(1-\delta)$, which is supposed not 
to change qualitatively the low energy physics. Hence, the 
effective model describing the squeezed chain consists of a dimerized 
Heisenberg chain with a domain wall. 
The model in the presence of a single domain wall can also be analyzed 
along the same lines outlined in 
Ref. \cite{Suzumura}. The result would be again that a variational 
solution can be obtained with a space dependent classical phase 
$\phi_c(x)$, but a space independent mass for the bosons describing 
the quantum fluctuations. The same would hold also in the XX-limit, which can 
be also easily solved by diagonalizing the spinless fermion Hamiltonian 
which is obtained through a Jordan-Wigner transformation. 
In the continuum limit and after linearizing the band around the Fermi 
momenta $k_F=\pm \pi/2$, one has to solve the following equations
\begin{eqnarray}
-i\frac{\partial}{\partial x}\chi_{\epsilon R}(x) 
- i \Delta(x) \chi_{\epsilon L}(x) &=& \epsilon \chi_{\epsilon R}(x), 
\label{chiR}\\
i\frac{\partial}{\partial x}\chi_{\epsilon L}(x) 
+ i \Delta(x) \chi_{\epsilon R}(x) &=& \epsilon \chi_{\epsilon L}(x), 
\label{chiL}
\end{eqnarray}
where the Fermi field is expressed in terms of the $\chi$ functions through
\[
\Psi(x) = \sum_\epsilon \left[
{\rm e}^{ik_F x}\chi_{\epsilon R}(x)    
+ {\rm e}^{-ik_F x}\chi_{\epsilon L}(x)\right] c_{\epsilon}
\equiv \sum_\epsilon \phi_\epsilon(x) c_{\epsilon}, 
\]
$c_\epsilon$ being the annihilation operator of an eigenstate of energy 
$\epsilon$. The dimerization parameter is equal to $\Delta$ for 
$x<0$, and to $-\Delta$ at $x>0$, thus having a domain wall at $x=0$.  
Eqs.(\ref{chiR}) and (\ref{chiL}) can be solved and one finds, in addition 
to scattering solutions with energy $\epsilon = \pm \sqrt{k^2+\Delta^2}$, 
$k$ being the wavevector, a localized zero energy solution with 
wavefunction
\[
\phi_0(x) = \sqrt{2\Delta}{\rm e}^{-|x|\Delta}\sin (k_Fx).
\]
This is a soliton solution for a single domain wall. 

The classical phase which appears in the bosonized version of the model 
can be related to the phase of the $2k_F$ oscillating part of the 
spinless fermion density. Namely, we expect that 
\begin{equation}
\rho_{2k_F}(x) = \langle \Psi^\dagger(x)\Psi(x) \rangle_{2K_F} 
\sim \cos\left(2 k_Fx + \phi_c(x)\right).
\label{phi-def}
\end{equation}
The dimer order corresponds to a bond ordered Charge Density Wave (CDW), 
i.e. $\phi_c(x)=(2n+1)\pi/2$, while the N\'eel phase to a site ordered 
CDW, i.e. $\phi_c(x)=n\pi$. In the absence of the domain wall, 
the phase is $\phi_c(x)=(2n+1)\pi/2$ at any $x$. In addition, in the 
presence of the domain wall, two other terms arise.      
One is  
\begin{equation}
\rho_1(x) = \left(n_0 - \frac{1}{2}\right)  \Delta {\rm e}^{-2|x|\Delta}
\left( 1 - \cos(2 k_F x)\right), 
\label{finale}
\end{equation}
where $n_0$ is the occupation probability of the soliton state. 
In fact, the term proportional to $n_0$ is the soliton wave function density 
probability, while the other one comes from the negative energy scattering solutions.
We notice that 
$\rho_1$ oscillates with a phase which has two possible values $\phi_c=0,\pi$ 
depending on $n_0$, which are compatible with the two possible 
N\'eel ordered phases. The other term which is generated by the domain wall 
decays like 
\begin{equation}
\rho_2(x) \sim \frac{\sin(2k_Fx)}{\sqrt{\Delta x}}{\rm e}^{-2|x|\Delta},
\label{rho2}
\end{equation}
and generates the polarization of the background responsible for the 
coupling among consecutive solitons, which we have considered throughout 
this work. This term oscillates with a phase compatible with a dimer order. 
Were $\rho_1(x)$ be present, we 
would indeed find a coexistence of dimer order and N\'eel one, the latter 
mainly localized close to the domain wall.    

We have now all the elements to describe what happens in this solvable model 
for the classical phase, and compare with the variational approach of 
Ref. \cite{Suzumura}. We start reminding that after the Jordan-Wigner 
transformation, a site occupied by a spinless fermion corresponds to 
a the spin having a $z$-component $S_z=1/2$, while an empty site to 
$S_z=-1/2$. Moreover, a chain with periodic boundary conditions is compatible 
with the presence of a single domain wall only if it has an odd 
number of sites. Therefore the ground state has a total $S=1/2$. In our 
XX-limit we can not rigorously discuss the SU(2) symmetry, but 
at least the following discussion gives some hints how the implementation 
of such symmetry gives a classical phase $\phi_c(x)=\pi/2$ independent of 
the position. In fact, if we take a ground state which does not break the 
symmetry, we have to assume that the chemical potential crosses exactly 
the soliton energy, so that there is an equal probability to have 
$S_z=\pm 1/2$. In other words this implies that $n_0=1/2$, leading to 
a vanishing $\rho_1(x)$, and leaving only the $\rho_2(x)$ term which 
has indeed a phase fixed at $\phi_c(x)=\pi/2$. 
Let us suppose a domain wall and an anti-domain wall 
at a distance much larger that $1/\Delta$. We can now consider a periodic 
chain with an even number of sites. The ground state will be a singlet. 
Around each domain wall there will be a localized soliton state at 
zero energy. The two $S=1/2$ solitons will be coupled by the polarization of 
the background, generating a singlet $S=0$ as well as a 
triplet $S=1$ state. The singlet state has energy $-t$ ($0<t\ll \Delta$), 
the $S_z=0$ component of the triplet $+t$, and the $S_z=\pm 1$ have energy 
zero, where $t$ decays exponentially with the distance among the 
two domain walls.    
The fact that the triplet is not degenerate is a consequence of the 
XX-limit. In this case, however, we are lucky since the lowest energy state, 
being a singlet, is SU(2) invariant. This is in essence why the spin 
anisotropy with a random distribution of domain walls does not play a very 
relevant role for what it regards the ground state properties. 
Within the lowest energy singlet state, each soliton state 
has an equal probability of being empty or occupied, leading again to a 
vanishing $\rho_1(x)$ Eq.(\ref{finale}). However, one can view 
the singlet state as oscillating between two configurations each having 
$\rho_1(x)\not = 0$, hence a classical phase $\phi_c=0,\pi$  
compatible with a N\'eel order.      
In a rough picture, the system 
is locally, in time, in one of the two possible N\'eel ordered states,  
and oscillates between them over a very long time scale $\sim 1/t$. 
Moreover, in spite of the exponential decay of the polarization over a
length $\xi_{SP}\sim 1/(2\Delta)$, which is the correlation length of the 
dimerized state, the two solitons, which are at a distance much larger that 
$\xi_{SP}$, oscillate coherently. This is on a very small scale our picture 
of what happens in the doped CuGeO$_3$. 

We think that the variational 
approach of Ref. \cite{Suzumura} is able to describe only the states 
where one soliton is occupied and the other is empty, or viceversa, and 
not the superposition of the two, which is beyond a mean field like 
approach. Therefore the method implicitly breaks SU(2) symmetry, which 
obviously leads to a finite staggered magnetization $\rho_1(x)$, since 
already the dimerized state breaks the translational invariance.  
For this reason, we believe that this approach 
should give a reasonable description of the 
N\'eel ordered phase, but is unable to describe the way in which the 
magnetic ordering is established, which is what we have done 
in this work.

\newpage


\begin{thebibliography}{99}
\bibitem{Cobalt} Isolated Co$^2+$ is in a spin 3/2 orbitally 
degenerate multiplet. Due to the orbital degeneracy, which is unclear 
if and how is split in CuGeO$_3$, it is still not possible to construct 
a realistic model for describing Co doping.
\bibitem{Fisher-note} This is the typical
behavior of a random Heisenberg chain 
as obtained by D. Fisher in Ref. \protect \cite{Fisher}.
\end{thebibliography}
\end{document}